\documentclass[bibyear]{aa}
\usepackage{graphicx}
\usepackage{txfonts}
\usepackage{xcolor}
\usepackage{comment}
\usepackage{natbib}
\usepackage{placeins} 
\usepackage{lipsum}

\titlerunning{star formation in supermassive black holes} 
\authorrunning{Chen et al.}
 
\begin{document} 

 \title{The relation between black hole spin, star formation rate, and black hole mass for supermassive black holes}

   \author{
        Yongyun Chen$\dagger$\inst{1} 
        \and
        Qiusheng Gu$\dagger$\inst{2}
        \and
        Junhui Fan\inst{3}
         \and
        Xiaotong Guo\inst{4}    
        \and
        Dingrong Xiong\inst{5}
       \and
        Xiaoling Yu\inst{1} 
        \and        
        Xiaogu Zhong\inst{1}
        \and
        Nan Ding\inst{6} 
        }

   \institute{
        $^{1}$ College of Physics and Electronic Engineering, Qujing Normal University, Qujing 655011, P.R. China\\
        $^{2}$ School of Astronomy and Space Science, Nanjing University, Nanjing 210093, P. R. China\\
        $^{3}$ Center for Astrophysics, Guangzhou University, Guangzhou 510006, China\\
        $^{4}$ Anqing Normal University, 246133, P. R. China\\
        $^{5}$ Yunnan Observatories, Chinese Academy of Sciences, Kunming 650011,China \\
        $^{6}$ School of Physical Science and Technology, Kunming University 650214, P. R. China\\
        \email{ynkmcyy@yeah.net;qsgu@nju.edu.cn}
             }

\abstract {Both theoretical models and observational evidence indicate that jets and/or outflows driven by central active supermassive black holes exert a significant feedback effect on the overall properties of their host galaxies. Theoretical models suggest that the spin of supermassive black holes drives relativistic jets. Therefore, we investigate the relationship between black hole spin, star formation rate, and black hole mass using a sample of 48 low-redshift supermassive black holes. By performing multiband fitting of spectral energy distribution, we derive the star formation rates and stellar masses of the host galaxies harbouring these supermassive black holes. Our main results are as follows: (i) For black holes with masses \(M_{\rm BH} \lesssim 10^{6.5} M_{\odot}\), the spin increases with increasing black hole mass, suggesting that black hole growth is primarily driven by gas accretion, particularly in the coherent gas accretion regime. Conversely, for black holes with masses \(M_{\rm BH} \gtrsim 10^{7.5} M_{\odot}\), the spin decreases with increasing black hole mass, indicating that growth occurs mainly through mergers, inducing chaotic accretion. (ii) At low star formation rates, black hole spin increases with increasing star formation rates, consistent with gas accretion. However, at high star formation rates, black hole spin decreases with increasing star formation rates, suggesting black hole mergers. The value of the black hole spin may be used to diagnose the star formation rate of the host galaxies through active galactic nuclei activities. (iii) Our data and analysis confirm the well-known relation between stellar mass and black hole mass, with the fitting function $\log M_{\rm BH}=0.57\log M_{*}+1.94$.}


 \maketitle
%
%-----------------------------------------------------------------

\section{Introduction}
It is widely accepted that galaxies possess a supermassive black hole at their centre \citep[e.g.][]{Kormendy1995, Magorrian1998, Gebhardt2000, Ferrarese2000, Tremaine2002, Marconi2003, DiMatteo2003a, DiMatteo2003b, Haring2004, Ho2008, DiMatteo2008, Gultekin2009, Greene2010, Alexander2012, Kormendy2013, Heckman2014}. The origin of these supermassive black holes has remained unclear; however, it seems relatively certain that the majority of black hole growth takes place through gas accretion \citep{Soltan1982}. The accretion process is accompanied by a significant release of energy, which is most notable in the emission of powerful quasars \citep{LyndenBell1969}. 

One of the significant enigmas regarding galaxy formation and evolution is the coevolution of supermassive black holes and host galaxies. The relatively close connection between the properties of the galaxy and the mass of the central supermassive black hole \citep[e.g.][]{Kormendy2001, Marconi2003, Ferrarese2006, Kormendy2013, Heckman2014} suggests that the black hole is not just a bystander to the formation process of the galaxy \cite[e.g.][]{DiMatteo2000, DiMatteo2001, DiMatteo2003a, DiMatteo2003b, DiMatteo2004, Di2005, DiMatteo2008, DiMatteo2012}, but affects it in a decisive way \citep{Bustamante2019}. This perspective has facilitated the incorporation of black hole evolution and feedback mechanisms into modern theories of galaxy formation \citep{Kauffmann2000, Bustamante2019}. There are two main ways in which active galactic nuclei (AGN) feedback can impact its host galaxy. The outflow of AGN can enhance star formation (positive feedback) by compressing molecular clouds and/or interstellar medium (ISM) \cite[e.g.][]{Croft2006, Rod2007, Feain2007, Elbaz2009, Imanishi2011, Crockett2012, Ishibashi2012, Zinn2013, Schaye2015, Schutte2022, Venturi2023}. Conversely, AGN can suppress star formation (negative feedback) through mechanical energy from wind, outflow, or jets, thereby heating the surrounding ISM and preventing molecular gas from radiatively cooling, or expelling gas from the host galaxy due to AGN-driven outflow \citep[e.g.][]{Cresci2015a, Cano2012, Carniani2016, Bischetti2022, Chen2022, Lammers2023}. Many simulation models explain the quenching transition of galaxies from star-forming blue clouds to the red sequence of elliptical galaxies due to black hole feedback \citep{Di2005, Springel2005, Hopkins2006, Sijacki2007, Croton2006, Dubois2013, Silk2013, Dubois2016, DeGraf2015, Hopkins2016}. Subsequently, some numerical simulations discovered that AGN triggers star formation \citep{Santini2012, Gaibler2012, Ishibashi2012, Liu2013, Zubovas2013}, which implies positive feedback. Therefore, supermassive black holes are a key element in any modern theory of galaxy formation and evolution \citep{Bustamante2019}. 

Some observations have demonstrated that the interaction between the radio jets and ISM of the host galaxy can also be accountable for large outflows \citep[e.g.][]{Emonts2005, Morganti2007, Morganti2013a, Morganti2013b, Tadhunter2014, Mahony2016}, and jet-driven outflows can exert a relevant role in feedback mechanisms. The theoretical model of jet formation indicates that the jet extracts rotational energy from the black hole and the jet power ($P_{\rm jet}$) depends on the spin of the black hole ($a$) \citep{blandford1977, Macdonald82, Thorne1986,  Ghisellini2006}, which implies that the spin of a black hole plays a crucial role in determining relativistic jets \citep{Livio1999, Meier2002, Koide2002, Garofalo2010, Tchekhovskoy12}. The black hole spin parameter ($0\leq |a|\leq1$) is $|a|=J_{\rm BH}c/GM_{\rm BH}^{2}$, where $M_{\rm BH}$ represents the black hole mass, $G$ is the gravitational constant, and $c$ is the vacuum speed of light. Observational evidence for the spin-enhanced jets has been reported for stellar mass black holes in X-ray binaries \citep{Narayan2012, Steiner2013}. Numerical simulations have also suggested that relativistic jets rely on the spin of black holes \citep[e.g.][]{Tanabe2008, McKinney2005}. Recently, \cite{Cui2023} have discovered the precessing jet nozzle connecting to a spinning black hole in M87. \cite{Unal2020} propose that the activity of a black hole depends on its spin. These aforementioned results indicate that the spin of a black hole can serve as a good indicator of the activity of a black hole. Thus, we investigated the relation between black hole spin and star formation.

In this article, we discuss how we primarily employed multiband fitting to acquire the stellar mass and star formation rate of the host galaxy of AGN and subsequently investigated the relationship between black hole spin, star formation rate, and black hole mass. The second section delineates the sample we used, the third section presents our results and discussion, and the fourth sections provides our conclusion.

\section{The sample}
\subsection{The sample of supermassive black holes}    
We collected a large sample of 48 low-redshift supermassive black holes with the spin of the black hole to study the relationship between black hole spin, star formation rate, and black hole mass. The redshift range of these supermassive black holes spans from 0.002 to 1.70. The redshift of the farthest supermassive black hole is 1.695. Hence, these supermassive black holes constituted a very local sample. We mainly focused on the sample from \cite{Reynolds2021}. The spin of the supermassive black hole was measured through X-ray reflection spectroscopy \citep[e.g.][]{Gallo2011, Walton2013, Risaliti2013, Lohfink2013, Brenneman2013, Reis2014, Keck2015, Parker2018, Sun2018, Buisson2018, Walton2019, Jiang2019, Walton2020}. The systematic uncertainties in X-ray reflection–based spin measurements are the uncertainties of spectral models. There are two distinct aspects to this. Firstly, the structure of the accretion disk was approximated and assumed when modelling the X-ray reflection spectrum. The second and more serious issue is whether we correctly attributed the observed spectral structure to inner disk reflections. Despite these uncertainties, black hole spin measurements using the X-ray reflection technique show statistical robustness \citep{Reynolds2021}. We also note that approximately 75\% of the sources in our sample have relatively high black hole spins, $a\geq0.8$. It is unlikely that the relativistic reflection method is biased towards measuring high black hole spin, as the spectral shape produced by low spin is less blurry and thus easier to distinguish from the underlying continuum \citep{Draghis2023}. Additionally, low and moderate spins are still detected by some authors \citep{Wang2018, Draghis2021, Jana2021, Jia2022}. However, one cannot exclude the possibility that there is an observational bias, which makes highly spinning black holes easier to detect. \cite{Vasudevan2016} explain that the black hole spin distribution measured in AGN is biased towards high values due to the selection effect of flux-limited observations \citep{Brenneman2011}. The black hole mass has been measured \citep[e.g.][]{Peterson2004, Walton2019, Bentz2006, Jiang2019, Parker2018, Nikolajuk2009, Walton2020, Sluse2012}. The mass of black holes is mainly estimated by using the virial method. The uncertainty of black hole mass is approximately 0.5 dex \cite[e.g.][]{Gebhardt2000, Ferrarese2001, Peterson2004}. We obtain the star formation rate and stellar mass of the host galaxy of a supermassive black hole by fitting multiband spectral energy distribution (SED) data. The sample is shown in Table A.1 of Appendix A.

\begin{table*}
	\centering
	\caption{\textsc{CIGALE} grid parameter values adopted for the modelling described.}
	\begin{tabular}{p{0.15\textwidth}p{0.25\textwidth}p{0.45\textwidth}}
		\hline
		\hline
		Parameter & Values & Description \\
		\hline
		\multicolumn{3}{c}{Star formation history (SFH): Delayed}\\
		$\tau_{\textnormal{main}}$&5–8000 (in steps of 10)& e-folding time of the main stellar population model (Myr).\\
		Age &200–13260 (in steps of 10)&Age of the oldest stars in the galaxy (Myr).\\
		\hline
		\multicolumn{3}{c}{Single-age stellar population (SSP): \cite{Bruzual2003}}\\
		imf & 1  &  Initial mass function \cite{Chabrier2003}\\
		Metallicity & 0.02 &  solar \\
		Separation Age& 10& Age of the separation (to differentiate) between the young and the old star populations (Myr).\\
		\hline
		\multicolumn{3}{c}{Dust attenuation:  \cite{Calzetti2000}}\\
		E$(B-V)_{\textnormal{young}}$ &0.015, 0.02, 0.04, 0.06, 0.1, 0.2, 0.3, 0.9, 1.0 & Color excess of the stellar continuum light for the young population.\\
		\hline
		\multicolumn{3}{c}{Dust emission: dl2014}\\
		Q$_{\rm pah}$ & 1.12, 1.77, 2.50, 3.19 & Mass fraction of PAH \\
		U$_{\rm min}$ & 5.0, 6.0, 7.0, 8.0, 10.0, 12.0, 15.0, 17.0, 20.0, 25.0 & Minimum radiation field \\
		A$_{\rm lpha}$ & 2.0, 2.1, 2.2, 2.3, 2.4, 2.5, 2.6, 2.7, 2.8 & the power-law distribution ($dM_{\rm d} \propto U^{-\alpha} dU$) \\
		Gamma     &  0.02 & Fraction illuminated from U$_{\rm min}$ to U$_{\rm max}$ \\
		\hline
		\multicolumn{3}{c}{AGN model: \cite{Fritz2006}}\\
		R$_{\textnormal{max}}$/R$_{\textnormal{min}}$ & 30.0, 60.0, 100.0, 150.0 & Ratio of the maximum to minimum radii of the dust torus. \\
		$\tau$ & 0.3, 0.6, 1.0, 2.0, 3.0, 6.0, 10.0& Optical depth at 9.7 $\mu$m. \\
		$\beta$ & $-$1.00, $-$0.75, $-$0.50, $-$0.25, 0.00 & Beta from the power-law density distribution for the radial component of the dust torus (eq. 3 of Fritz 2006).\\
		$\gamma$ & 0.0, 2.0, 4.0, 6.0 & Gamma from the power-law density distribution for the polar component of the dust torus (eq. 3 of \cite{Fritz2006}).\\
		Opening Angle ($\theta$) & 60.0, 100.0, 140.0 & Full opening angle of the dust torus (Fig 1 of Fritz 2006). \\
		$\psi$ & 0.001, 10.1, 20.1, 30.1, 70.1, 89.990 & Angle between equatorial axis and line of sight. \\
		$f_{\rm{AGN}}$ &  0.0, 0.05, 0.1, 0.15, 0.2, 0.25, 0.3, 0.35, 0.4, 0.45, 0.5, 0.55, 0.6, 0.65, 0.7, 0.75,
		0.8, 0.85, 0.9, 0.95, 0.99 & Fraction of AGN torus contribution to the IR luminosity (fracAGN in Eq. 1 of \cite{Ciesla2015})\\
		\hline
	\end{tabular}
	\label{tab:Par_CIG}
\end{table*}

\subsection{The stellar masses and star formation rates of AGN host galaxies}
\begin{figure}
	\centering
	\begin{tabular}{@{\extracolsep{\fill}}c@{}c@{\extracolsep{\fill}}}
		\includegraphics[width=\linewidth]{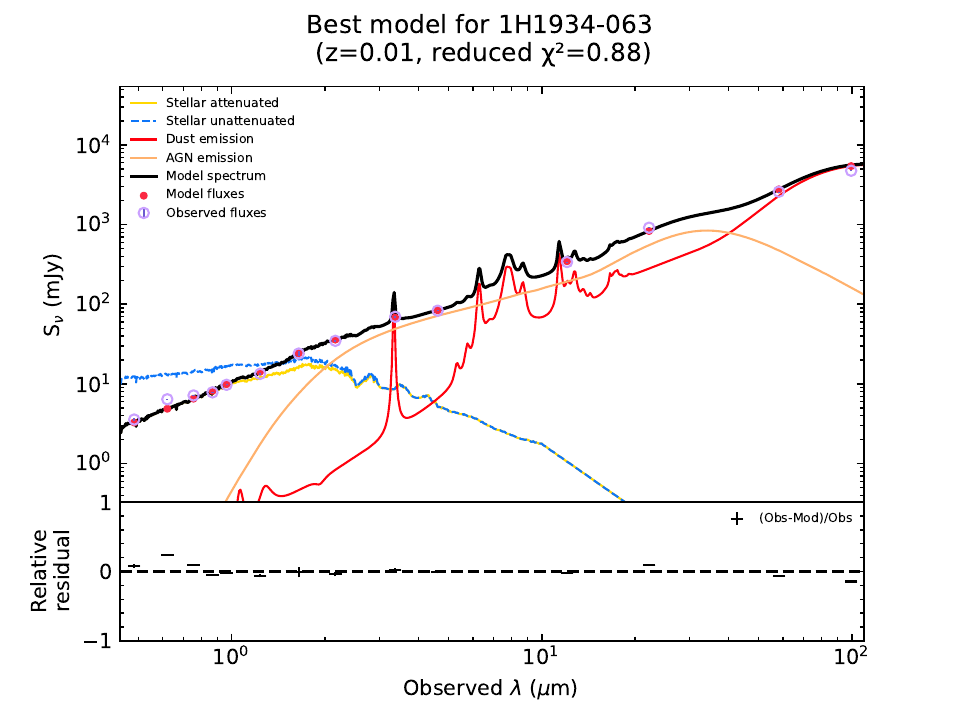}
		%	(a)OpenCV的logo & (b)GDAL的logo\\
	\end{tabular}
	\caption{
		Example of CIGALE SED fitting for our sample.
		The black line indicates the best-fitting model. The blue, orange, red, and yellow lines represent unattenuated stellar, attenuated stellar, dust, and AGN emission, respectively. The lower panel indicates the residual of the best fitting.
	}
	\label{SEDmodel}
\end{figure}

We derived the stellar masses and star formation rates of the host galaxies by using version 2022.0 of the SED-fitting code {\bf CIGALE} \citep{Burgarella2005, Noll2009, Boquien2019}. We adopted the optical photometry data from Pan-STARRS1-DR1 Surveys \citep{Chambers2016, Magnier2020}. The IR data were obtained from the Two Micron All Sky Survey (2MASS) \citep{Skrutskie2006} and Wide-field Infrared Survey Explorer (WISE) \citep{Wright2010}. The far-infrared photometric data from IRAS (60 and 100 $\mu$m) were referred to in \cite{Beichman1988}. As a result, we compiled a total of 14 photometric data (g, r, i, z, y$_{P1}$, J, H, and K/Ks; as well as at 3.4, 4.6, 12, 22, 60, and 100 $\mu$m) for the SED fitting. In our work, we utilized the templates of galaxy plus AGN to fit the SEDs of our sample. The template for galaxies consists of four modules, including star formation history (SFH), single stellar population model \citep{Bruzual2003}, dust attenuation \citep{Calzetti2000}, and dust emission \citep{Draine2007, Dale2014}. The module employed for the component of AGNs was \cite{Fritz2006}. All these modules are included in CIGALE. We defined the CIGALE grid of model galaxy SEDs using the parameters and values given in Table 1. CIGALE identifies the best-fit SED model by minimizing the $\chi^{2}$ statistic. We present examples of the best SED fitting in Figure.\ref{SEDmodel}. We also note that the star formation rate estimated by the SED fitting is more dependent on the star formation history
model. Therefore, we used the calibration from \cite{Kennicut1998} to estimate the star formation rate by the IR luminosities. The IR luminosity of the host was calculated by integrating the host galaxy far-IR SED subcomponent corresponding to the cold dust emission. The star formation rate was calculated by using the formula established by \cite{Xie2021}, renormalized to a \cite{Chabrier2003} initial mass function (IMF):

\begin{equation}
	\rm SFR(M_{\odot} yr^{-1}) = 4.5\times10^{-44}L_{IR}(erg~s^{-1}), 
\end{equation}
where $L_{\rm IR}$ is the $8-1000~\mu \rm m$ rest-frame IR luminosity. Some authors have also derived the star formation rate by using the relation from \cite{Kennicut1998} for different IMF \citep[e.g.][]{Molina2023, Suh2019}, such as \cite{Kroupa2001} and \cite{Chabrier2003} IMFs. The differences in star formation rates that are obtained using different IMFs are minimal, typically not exceeding an order of magnitude.  Following \cite{Xie2021}, we also used equation (1) to calculate the star formation rate of the AGN host galaxy, because the \cite{Chabrier2003} IMF was assumed when fitting the SED for our sample. \cite{Xie2021} find that the true uncertainties in L$_{\rm IR}$ are approximately $\sim$0.2–0.3 dex. Thereafter, we assumed this scatter value as the 1$\sigma$ error of our star formation rate estimates.

\begin{figure}
	\includegraphics[width=\linewidth]{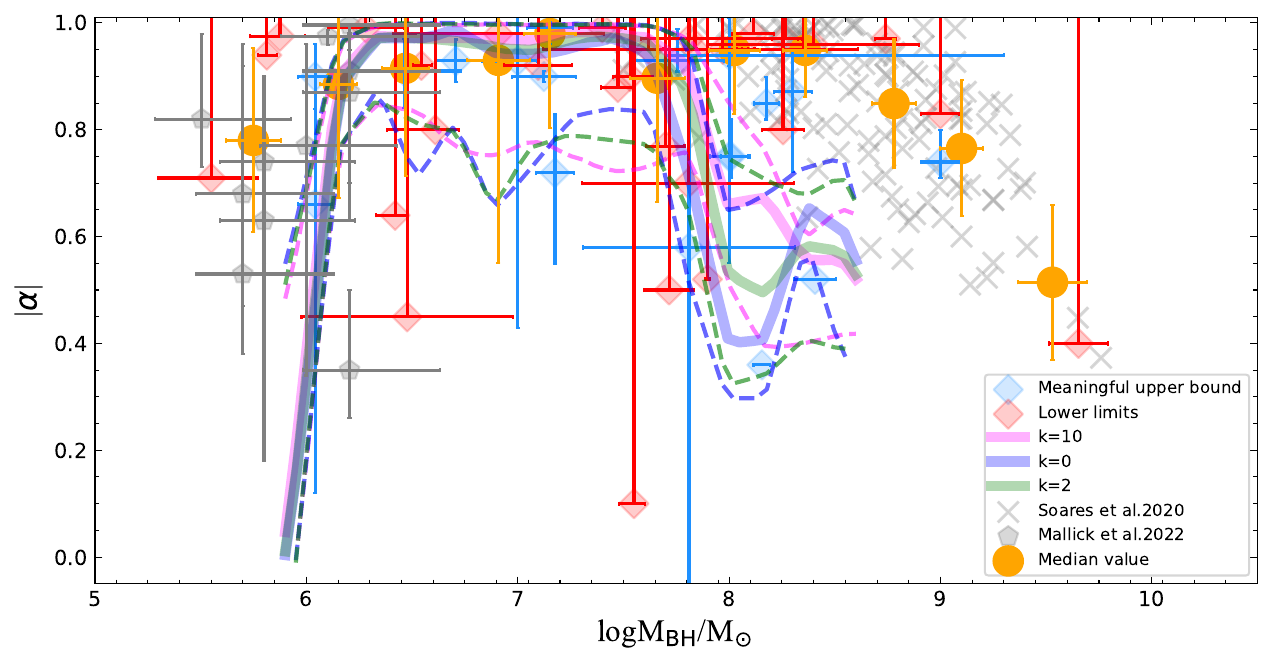}
	\centering
	\caption{Relation between black hole spin and black hole mass for a supermassive black hole. The red diamonds are the lower limits and the blue diamonds are measurements that include a meaningful upper bound. The error bars in mass are the 1$\sigma$ errors from Table 1; where that was not available, we assumed a 0.5 dex \citep{Reines2015, Koss2022, Reynolds2021}. The solid lines are the black hole spin evolution model of \cite{Bustamante2019}. The dashed lines represent 25 and 75 percent percentiles. For completely chaotic accretion ($k=0$), the black hole spin no longer follows the angular momentum of the gas. For a very coherent accretion ($k = 10$), the black hole behaves similarly to during the coherent regime. Medium concentrations (e.g. $k=2$) produce a slightly intermediate behaviour, where the black hole is fairly aligned with the angular momentum of the gas, but occasionally counterrotating accretion episodes still occur. The grey pentagons are observations from \cite{Mallick2022} and the grey crosses are from \cite{Soares2020}. The orange circles are the median values of black hole spin in each bin.}
	\label{spinMBH}
\end{figure}

\section{Results and Discussions}
\subsection{Relation between black hole mass and black hole spin}
Constraining the primary growth mechanisms of supermassive black holes remains one of the most actively debated topics in cosmological structure formation. Given the anticipated relationship between the evolution of supermassive black hole spin parameters and the accretion and merger histories of individual black holes \citep{Shakura1976, Berti2008, Barausse2012}, black hole spin may offer valuable insights into the cosmic growth of supermassive black holes.

To facilitate a more comprehensive comparison between the observed data and the black hole spin evolution model that \cite{Bustamante2019} propose, we incorporated additional samples from \cite{Mallick2022} and \cite{Soares2020}. Specifically, the dataset from \cite{Mallick2022} contributed information on the spin of a low-mass black hole, while the sample from \cite{Soares2020} pertains to the spin of a massive black hole. We also note that the black hole spin that \cite{Soares2020} report is not directly measured via the X-ray reflection method; instead, it is inferred based on jet production efficiency. Due to the absence of multiband data, such as far-infrared observations, for the samples from \cite{Mallick2022} and \cite{Soares2020}, we did not fit multiband data to obtain star formation rates and stellar masses. Figure.\ref{spinMBH} shows the relation between black hole spin and black hole mass for our sample. The observed data is similar to the black hole spin evolution model of \cite{Bustamante2019}. For black holes with masses less than 10$^{6.5}$M$_{\odot}$, the spin of a black hole increases with the mass of the black hole, which implies that the primary mechanism driving black hole growth in this mass range is gas accretion. This is consistent with the spin evolution model of \cite{Bustamante2019}, \cite{Dubois2021}, and \cite{Sala2024}. In the beginning, the spin evolution of a black hole mainly depends on the black hole mass, no matter how fast the mass accretes. The spin of a low-mass black hole can be reoriented during an accretion episode, making counterrotating accretion highly improbable and leading to continuous spin-up \citep{Bustamante2019}. Conversely, for black holes exceeding 10$^{7.5}$M$_{\odot}$, the spin tends to decrease to intermediate levels. This phenomenon is attributed to black hole mergers and the transition into the self-gravity regime, where the frequency of counterrotating accretion episodes is governed by the concentration parameter \citep{Bustamante2019}.  \cite{Pacucci2020} adopt a theoretical framework to assess the relative contributions of accretion and mergers to black hole growth across different mass scales and redshifts. Their findings indicate that, for black holes with masses greater than 10$^{8.0}$M$_{\odot}$, mergers become the dominant growth mechanism, resulting in a decline in black hole spin. At this mass threshold, kinetic feedback mechanisms become significant and black hole coalescence emerges as the most efficient channel for spin evolution, surpassing the thermal quasar mode, which restricts gas accretion \citep{Weinberger2017}. We also find that the black hole spin tends to have very high values for intermediate black hole mass, but decreases for more massive black holes. Our results are also consistent with previous numerical studies of spin evolution in a cosmological context \citep[e.g.][]{Dubois2021, Sala2024} and observational trends \citep[e.g.][]{Reynolds2021, Mallick2022}. According to \cite{Volonteri2005}, high spins in low-mass supermassive black hole systems suggest that these systems primarily grow through coherent accretion events. In contrast, more massive supermassive black holes ($M_{\rm BH}>3\times10^{7}M_{\odot}$) with modest spins imply that their growth is predominantly driven by supermassive black holes merging. The mass dependence of black hole spins can be also seen in semianalytical models \citep[e.g.][]{Berti2008, Fanidakis2011, Barausse2012, Dotti2013, Volonteri2013, Griffin2019, Izquierdo2020, Sesana2014, Zhang2019} and hydrodynamical simulations of galaxy formation \citep[e.g.][]{Dubois2014, Beckmann2024, Peirani2024}. 

\begin{figure}
	\includegraphics[width=\linewidth]{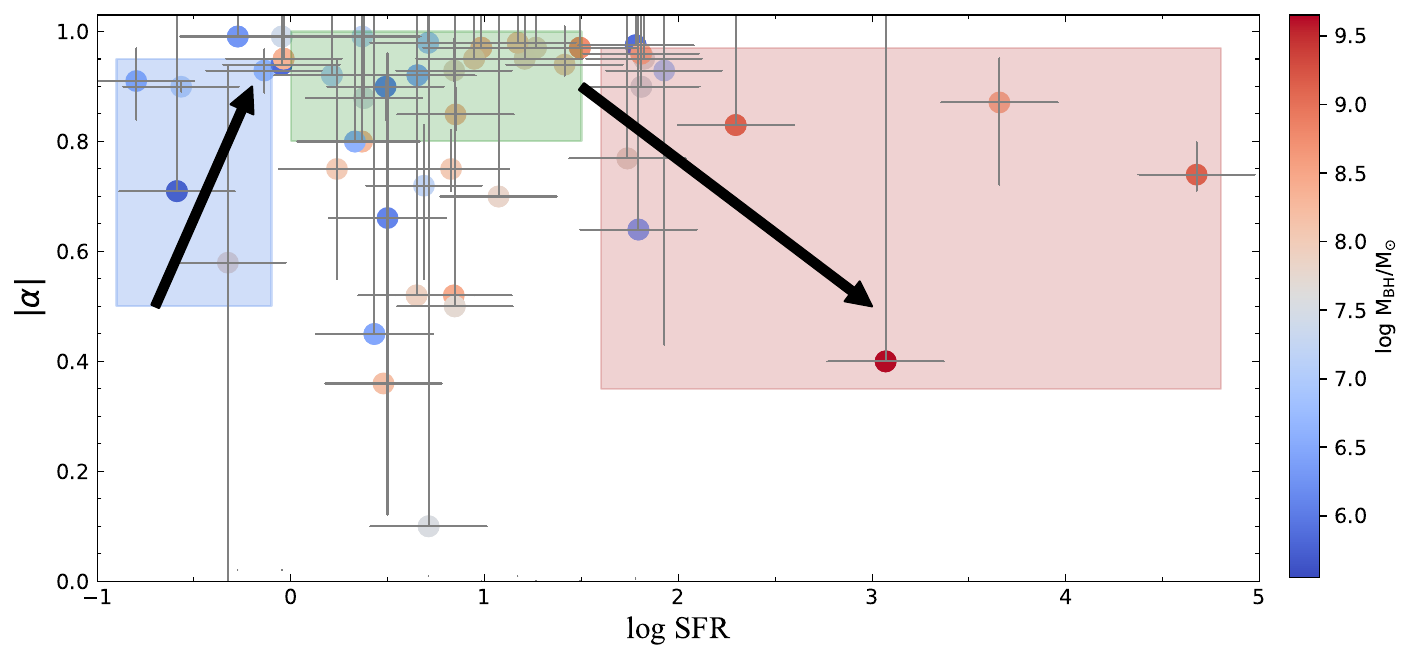}
	\centering
	\caption{Relation between the spin of the black hole and star formation rate for a supermassive black hole. All spin measurements reported here are from the X-ray reflection method. The colourbar indicates the black hole mass. The shaded areas represent low star formation rates, medium star formation rates, and high star formation rates, respectively.}
	\label{sfrspin}
\end{figure}

\subsection{Relation between star formation rate and black hole spin}
Figure. \ref{sfrspin} shows the relationship between the star formation rate and the spin of supermassive black holes. From Figure.\ref{sfrspin}, we find that the star formation rate increases with the increase of the spin of a black hole for the sources with low star formation rates (log SFR$<0~\rm M_{\odot} yr^{-1}$). The possible explanation for the sources with low star formation rates is gas accretion. The sources with low star formation rates tend to have low black hole mass. According to Figure.\ref{spinMBH}, we find that the growth of low-mass black holes is predominantly attributed to gas accretion. The star formation rate increases as the spin of the black hole decreases for the sources with high star formation rate (log SFR$>1.5~\rm M_{\odot} yr^{-1}$). A possible explanation for the sources with high star formation rates is the merger of black holes, which is often associated with higher black hole masses. \cite{Dubois2014} find that the growth of massive black holes may be mainly due to the merger of black holes (their Figure 4). The black hole mergers may imply galaxy mergers. The galaxy mergers are assumed to enhance star formation \citep[e.g.][]{Barnes2004, Springel2005, Hopkins2006, Randall2008, Lin2008, Kim2009, Saitoh2009, Rupke2010, Perez2011, Ellison2013, Knapen2015, Athanassoula2016, Silva2018, Wolter2023}. For the sources with intermediate star formation rate ($\rm M_{\odot} yr^{-1}~0<\log \rm SFR<1.5~\rm M_{\odot} yr^{-1}$), we find that these sources have both high black hole spin and low black hole spin. Within this intermediate star formation rate range, where black hole masses are varied, both high and low black hole masses are present. This diversity may be caused by the combination of the merger of black holes and gas accretion, which can abruptly modify both spin magnitude and orientation \citep{Dubois2021}. 

The theory of jet formation \citep{blandford1977} holds that the spin of the supermassive black hole enhances the relativistic jet. This hypothesis has been supported by observational evidence indicating that black hole spin enhances the relativistic jet \citep{Narayan2012, Steiner2013}. \cite{Unal2020} find that black hole activity depends on the spin of a black hole. The amplitude and orientation of the black hole spin control the efficiency of the conversion of the gas accreted onto the black hole into energy \citep{Bustamante2019, Sala2024}. Consequently, black hole spin is a critical parameter influencing jet efficiency and AGN feedback, which in turn can modulate and self-regulate black hole accretion \citep{Peirani2024}. Both theoretical models \citep{Silk2009} and observations \cite[e.g.][]{Croft2006, Rod2007, Feain2007, Elbaz2009, Imanishi2011, Crockett2012, Zinn2013, Cresci2015a, Cano2012, Carniani2016, Bischetti2022, Chen2022, Schutte2022, Lammers2023, Venturi2023} have shown that the activity of AGN impacts the star formation of host galaxies. Our above results may imply that the activity of AGN provides a feedback effect on galaxies. By considering radiation feedback, \cite{Ishibashi2019} investigated the influence of the spin of the central black hole on the large-scale properties of the host galaxy. Black hole spin can generate significant macroscopic effects at galactic scales, ultimately leading to large-scale feedback. The value of the black hole spin may give a diagnostic about the star formation rate of the host galaxies through the activity of AGN, which is complex. However, we also note that measuring black hole spin is more challenging compared to measuring star formation rate. We also note that, due to observation limitations, black hole spin measurements for most of our samples obtained by X-ray reflection spectroscopy are greater than 0.5, which may lead to a selective effect. In the future, our above results should be tested using large samples.

\subsection{Relation between black hole mass and stellar mass}
\begin{figure}
	\includegraphics[width=8.0cm,height=8.0cm]{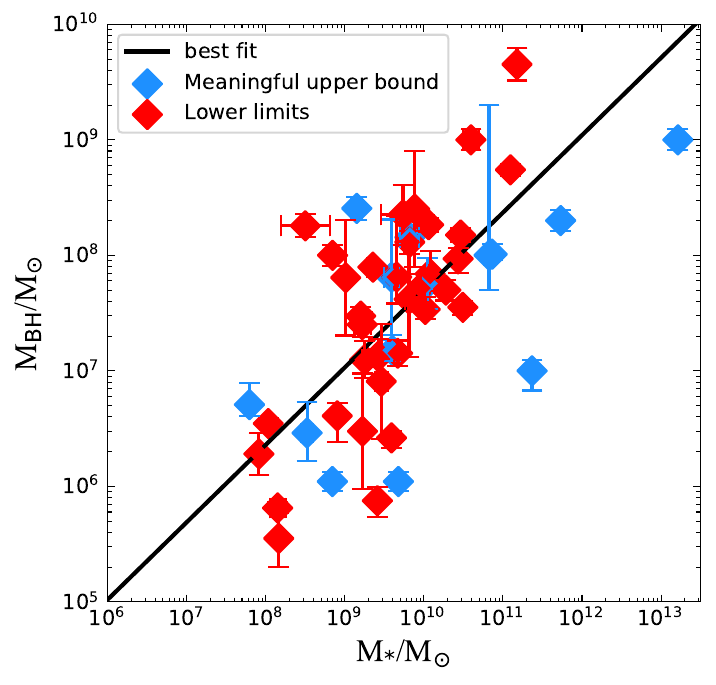}
	\centering
	\caption{
		Relation between the mass of the central supermassive black hole and stellar mass.
		The black line is the best fitting. The red diamonds are the lower limits and the blue diamonds are measurements that include a meaningful upper bound. 
	}
	\label{MbhMstar}
\end{figure}

The observed correlation between the mass of the central supermassive black holes of local galaxies and their stellar mass \citep{McLure2001, McLure2002, Kormendy2001, Marconi2003, Ferrarese2006, Kormendy2013} may imply the influence of AGN feedback on the star formation of the host galaxy. Investigating this mechanism is a primary objective in both observational and theoretical astronomy, as it could provide a crucial factor in our comprehensive understanding of galaxy formation. AGN or quasar feedback has often been invoked to explain the origin of coevolution of galaxies and supermassive black holes in the local universe \citep{Fabian2012}. Thus, we study the relation between the mass of central supermassive black holes and their stellar mass. We used  {\sffamily linmix} \citep{Kelly2007} to fit the relation between two variables of AGN. Figure.\ref{MbhMstar} shows the relation between the mass of central supermassive black holes and their stellar mass for our sample. We find that there is a strong correlation between the mass of central supermassive black holes and their stellar mass ($r=0.65, p<0.0001$; $r$ and $p$ are the correlation coefficient and probability for the null hypothesis of no correlation, respectively; the $p<0.05$ is a significant correlation),

\begin{equation}
	\log M_{\rm BH}=0.57(\pm0.11)\log M_{*}+ 1.94(\pm1.05),
\end{equation}
with an intrinsic scatter of 0.54 dex. We find that the slope of the relation between black hole mass and stellar mass for our sample is $0.57\pm0.11$. \cite{Beifiori2012} reported that the slope of the relation between black hole mass and bulge stellar mass is 0.79$\pm$0.26 using 46 galaxies. Our results are similar to theirs within error. \cite{Pacucci2023} find that the relation between black hole mass and stellar mass for high redshift active galaxies is $\log M_{\rm BH}=1.06(\pm0.09)\log M_{*}-2.43(\pm0.83)$, with an intrinsic scatter of 0.23 dex. Our results are inconsistent with theirs, which may be due to the lower redshift of our sample compared to theirs. \cite{Pacucci2023} also suggest that the standard local universe relation is no longer valid in the high-redshift universe, particularly for faint AGN. \cite{Reines2015} find that the relation between black hole mass and stellar mass for a sample of 262 broad-line AGN in the local universe is $\log M_{\rm BH}=1.05(\pm0.11)\log M_{*}/10^{11}+ 7.45(\pm0.08)$, with an intrinsic scatter of 0.24 dex. Our result is also different from the result of \cite{Reines2015}. The possible explanation for this difference is that our sample size was smaller than that of \cite{Reines2015}.

\section{Conclusions}
For this study, we primarily examined the relationships between black hole spin, star formation rate, black hole mass, and stellar mass utilizing a sample of 48 supermassive black holes. The key findings of our research are summarized as follows:

(1) We find that in black holes with a mass $M_{\rm BH}\lesssim10^{6.5}M_{\odot}$, the black hole spin increases with the black hole mass. However, the spin of the black hole decreases with the black hole mass when the latter is $M_{\rm BH}\gtrsim10^{7.5}M_{\odot}$. These results may suggest that the growth of black holes from sources with low black hole mass mainly comes from coherent gas accretion events, while the growth of black holes from sources with high black hole mass mainly comes from merging events, including chaotic gas accretion.

(2) For the sources with low star formation rate (log SFR$<0.0~\rm M_{\odot} yr^{-1}$), the star formation rate increases with the spin of the black hole, which implies gas accretion. For the sources with intermediate star formation rate ($0.0~\rm M_{\odot} yr^{-1}<log SFR<1.5~\rm M_{\odot} yr^{-1}$), more frequent physical processes, such as the combination of black hole merger and gas accretion, are implied. For the sources with a high star formation rate (log SFR$>1.5~\rm M_{\odot} yr^{-1}$), the star formation rate increases as the spin of the black hole decreases, which implies the merger of a black hole. The value of the black hole spin may give a diagnostic about the star formation rate of the host galaxies through AGN activities, which are complex.

(3) There is a significant correlation between black hole mass and stellar mass for supermassive black holes. Our best-fitting power-law relation for the full galaxy sample is $\log M_{\rm BH}=0.57\log M_{*}+1.94$. 

\begin{acknowledgements}
Yongyun Chen is grateful for financial support from the National Natural Science Foundation of China (No. 12203028). Yongyun Chen is grateful for funding for the training Program for talents in Xingdian, Yunnan Province (2081450001). 
QSGU is supported by the National Natural Science Foundation of China (12121003, 12192220, and 12192222).
We also acknowledge the science research grants from the China Manned Space Project with NO. CMS-CSST-2021-A05. This work is supported by the National Natural Science Foundation of China (11733001, U2031201 and 12433004). Xiaotong Guo acknowledge the support of \emph{National Nature Science Foundation of China} (Nos  12303017). This work is also supported by \emph{Anhui Provincial Natural Science Foundation} project number 2308085QA33. D.R.X. is supported by the NSFC 12473020, Yunnan Province Youth Top Talent Project (YNWR-QNBJ-2020-116) and the CAS Light of West China Program.
\end{acknowledgements}

\bibliographystyle{aa}
\bibliography{example} % if your bibtex file is called example.bib

\begin{thebibliography}{170}
\expandafter\ifx\csname natexlab\endcsname\relax\def\natexlab#1{#1}\fi

\bibitem[{{Afanasiev} {et~al.}(2019){Afanasiev}, {Popovi{\'c}}, \&
  {Shapovalova}}]{Afanasiev2019}
{Afanasiev}, V.~L., {Popovi{\'c}}, L.~{\v{C}}., \& {Shapovalova}, A.~I. 2019,
  \mnras, 482, 4985

\bibitem[{{Alexander} \& {Hickox}(2012)}]{Alexander2012}
{Alexander}, D.~M. \& {Hickox}, R.~C. 2012, \nar, 56, 93

\bibitem[{{Athanassoula} {et~al.}(2016){Athanassoula}, {Rodionov}, {Peschken},
  \& {Lambert}}]{Athanassoula2016}
{Athanassoula}, E., {Rodionov}, S.~A., {Peschken}, N., \& {Lambert}, J.~C.
  2016, \apj, 821, 90

\bibitem[{{Bambi} {et~al.}(2021){Bambi}, {Brenneman}, {Dauser}, {Garc{\'\i}a},
  {Grinberg}, {Ingram}, {Jiang}, {Liu}, {Lohfink}, {Marinucci}, {Mastroserio},
  {Middei}, {Nampalliwar}, {Nied{\'z}wiecki}, {Steiner}, {Tripathi}, \&
  {Zdziarski}}]{Bambi2021}
{Bambi}, C., {Brenneman}, L.~W., {Dauser}, T., {et~al.} 2021, \ssr, 217, 65

\bibitem[{{Barausse}(2012)}]{Barausse2012}
{Barausse}, E. 2012, \mnras, 423, 2533

\bibitem[{{Barnes}(2004)}]{Barnes2004}
{Barnes}, J.~E. 2004, \mnras, 350, 798

\bibitem[{{Beckmann} {et~al.}(2024){Beckmann}, {Smethurst}, {Simmons}, {Coil},
  {Dubois}, {Garland}, {Lintott}, {Martin}, {Peirani}, \&
  {Pichon}}]{Beckmann2024}
{Beckmann}, R.~S., {Smethurst}, R.~J., {Simmons}, B.~D., {et~al.} 2024, \mnras,
  527, 10867

\bibitem[{{Beichman} {et~al.}(1988){Beichman}, {Neugebauer}, {Habing}, {Clegg},
  \& {Chester}}]{Beichman1988}
{Beichman}, C.~A., {Neugebauer}, G., {Habing}, H.~J., {Clegg}, P.~E., \&
  {Chester}, T.~J., eds. 1988, {Infrared Astronomical Satellite (IRAS) Catalogs
  and Atlases.Volume 1: Explanatory Supplement.}, Vol.~1

\bibitem[{{Beifiori} {et~al.}(2012){Beifiori}, {Courteau}, {Corsini}, \&
  {Zhu}}]{Beifiori2012}
{Beifiori}, A., {Courteau}, S., {Corsini}, E.~M., \& {Zhu}, Y. 2012, \mnras,
  419, 2497

\bibitem[{{Bentz} {et~al.}(2006){Bentz}, {Denney}, {Cackett}, {Dietrich},
  {Fogel}, {Ghosh}, {Horne}, {Kuehn}, {Minezaki}, {Onken}, {Peterson}, {Pogge},
  {Pronik}, {Richstone}, {Sergeev}, {Vestergaard}, {Walker}, \&
  {Yoshii}}]{Bentz2006}
{Bentz}, M.~C., {Denney}, K.~D., {Cackett}, E.~M., {et~al.} 2006, \apj, 651,
  775

\bibitem[{{Berti} \& {Volonteri}(2008)}]{Berti2008}
{Berti}, E. \& {Volonteri}, M. 2008, \apj, 684, 822

\bibitem[{{Bischetti} {et~al.}(2022){Bischetti}, {Feruglio}, {D'Odorico},
  {Arav}, {Ba{\~n}ados}, {Becker}, {Bosman}, {Carniani}, {Cristiani}, {Cupani},
  {Davies}, {Eilers}, {Farina}, {Ferrara}, {Maiolino}, {Mazzucchelli},
  {Mesinger}, {Meyer}, {Onoue}, {Piconcelli}, {Ryan-Weber}, {Schindler},
  {Wang}, {Yang}, {Zhu}, \& {Fiore}}]{Bischetti2022}
{Bischetti}, M., {Feruglio}, C., {D'Odorico}, V., {et~al.} 2022, \nat, 605, 244

\bibitem[{{Blandford} \& {Znajek}(1977)}]{blandford1977}
{Blandford}, R.~D. \& {Znajek}, R.~L. 1977, \mnras, 179, 433

\bibitem[{{Boquien} {et~al.}(2019){Boquien}, {Burgarella}, {Roehlly}, {Buat},
  {Ciesla}, {Corre}, {Inoue}, \& {Salas}}]{Boquien2019}
{Boquien}, M., {Burgarella}, D., {Roehlly}, Y., {et~al.} 2019, \aap, 622, A103

\bibitem[{{Brenneman}(2013)}]{Brenneman2013}
{Brenneman}, L. 2013, {Measuring the Angular Momentum of Supermassive Black
  Holes}

\bibitem[{{Brenneman} {et~al.}(2011){Brenneman}, {Reynolds}, {Nowak}, {Reis},
  {Trippe}, {Fabian}, {Iwasawa}, {Lee}, {Miller}, {Mushotzky}, {Nandra}, \&
  {Volonteri}}]{Brenneman2011}
{Brenneman}, L.~W., {Reynolds}, C.~S., {Nowak}, M.~A., {et~al.} 2011, \apj,
  736, 103

\bibitem[{{Bruzual} \& {Charlot}(2003)}]{Bruzual2003}
{Bruzual}, G. \& {Charlot}, S. 2003, \mnras, 344, 1000

\bibitem[{{Buisson} {et~al.}(2018){Buisson}, {Parker}, {Kara}, {Vasudevan},
  {Lohfink}, {Pinto}, {Fabian}, {Ballantyne}, {Boggs}, {Christensen}, {Craig},
  {Farrah}, {Hailey}, {Harrison}, {Ricci}, {Stern}, {Walton}, \&
  {Zhang}}]{Buisson2018}
{Buisson}, D.~J.~K., {Parker}, M.~L., {Kara}, E., {et~al.} 2018, \mnras, 480,
  3689

\bibitem[{{Burgarella} {et~al.}(2005){Burgarella}, {Buat}, \&
  {Iglesias-P{\'a}ramo}}]{Burgarella2005}
{Burgarella}, D., {Buat}, V., \& {Iglesias-P{\'a}ramo}, J. 2005, \mnras, 360,
  1413

\bibitem[{{Bustamante} \& {Springel}(2019)}]{Bustamante2019}
{Bustamante}, S. \& {Springel}, V. 2019, \mnras, 490, 4133

\bibitem[{{Calzetti} {et~al.}(2000){Calzetti}, {Armus}, {Bohlin}, {Kinney},
  {Koornneef}, \& {Storchi-Bergmann}}]{Calzetti2000}
{Calzetti}, D., {Armus}, L., {Bohlin}, R.~C., {et~al.} 2000, \apj, 533, 682

\bibitem[{{Cano-D{\'\i}az} {et~al.}(2012){Cano-D{\'\i}az}, {Maiolino},
  {Marconi}, {Netzer}, {Shemmer}, \& {Cresci}}]{Cano2012}
{Cano-D{\'\i}az}, M., {Maiolino}, R., {Marconi}, A., {et~al.} 2012, \aap, 537,
  L8

\bibitem[{{Carniani} {et~al.}(2016){Carniani}, {Marconi}, {Maiolino},
  {Balmaverde}, {Brusa}, {Cano-D{\'\i}az}, {Cicone}, {Comastri}, {Cresci},
  {Fiore}, {Feruglio}, {La Franca}, {Mainieri}, {Mannucci}, {Nagao}, {Netzer},
  {Piconcelli}, {Risaliti}, {Schneider}, \& {Shemmer}}]{Carniani2016}
{Carniani}, S., {Marconi}, A., {Maiolino}, R., {et~al.} 2016, \aap, 591, A28

\bibitem[{{Chabrier}(2003)}]{Chabrier2003}
{Chabrier}, G. 2003, \pasp, 115, 763

\bibitem[{{Chambers} {et~al.}(2016){Chambers}, {Magnier}, {Metcalfe},
  {Flewelling}, {Huber}, {Waters}, {Denneau}, {Draper}, {Farrow}, {Finkbeiner},
  {Holmberg}, {Koppenhoefer}, {Price}, {Rest}, {Saglia}, {Schlafly}, {Smartt},
  {Sweeney}, {Wainscoat}, {Burgett}, {Chastel}, {Grav}, {Heasley}, {Hodapp},
  {Jedicke}, {Kaiser}, {Kudritzki}, {Luppino}, {Lupton}, {Monet}, {Morgan},
  {Onaka}, {Shiao}, {Stubbs}, {Tonry}, {White}, {Ba{\~n}ados}, {Bell},
  {Bender}, {Bernard}, {Boegner}, {Boffi}, {Botticella}, {Calamida},
  {Casertano}, {Chen}, {Chen}, {Cole}, {Deacon}, {Frenk}, {Fitzsimmons},
  {Gezari}, {Gibbs}, {Goessl}, {Goggia}, {Gourgue}, {Goldman}, {Grant},
  {Grebel}, {Hambly}, {Hasinger}, {Heavens}, {Heckman}, {Henderson}, {Henning},
  {Holman}, {Hopp}, {Ip}, {Isani}, {Jackson}, {Keyes}, {Koekemoer}, {Kotak},
  {Le}, {Liska}, {Long}, {Lucey}, {Liu}, {Martin}, {Masci}, {McLean}, {Mindel},
  {Misra}, {Morganson}, {Murphy}, {Obaika}, {Narayan}, {Nieto-Santisteban},
  {Norberg}, {Peacock}, {Pier}, {Postman}, {Primak}, {Rae}, {Rai}, {Riess},
  {Riffeser}, {Rix}, {R{\"o}ser}, {Russel}, {Rutz}, {Schilbach}, {Schultz},
  {Scolnic}, {Strolger}, {Szalay}, {Seitz}, {Small}, {Smith}, {Soderblom},
  {Taylor}, {Thomson}, {Taylor}, {Thakar}, {Thiel}, {Thilker}, {Unger},
  {Urata}, {Valenti}, {Wagner}, {Walder}, {Walter}, {Watters}, {Werner},
  {Wood-Vasey}, \& {Wyse}}]{Chambers2016}
{Chambers}, K.~C., {Magnier}, E.~A., {Metcalfe}, N., {et~al.} 2016, arXiv
  e-prints, arXiv:1612.05560

\bibitem[{{Chen} {et~al.}(2022){Chen}, {He}, {Ho}, {Gu}, {Wang}, {Zhuang},
  {Liu}, \& {Wang}}]{Chen2022}
{Chen}, Z., {He}, Z., {Ho}, L.~C., {et~al.} 2022, Nature Astronomy, 6, 339

\bibitem[{{Ciesla} {et~al.}(2015){Ciesla}, {Charmandaris}, {Georgakakis},
  {Bernhard}, {Mitchell}, {Buat}, {Elbaz}, {LeFloc'h}, {Lacey}, {Magdis}, \&
  {Xilouris}}]{Ciesla2015}
{Ciesla}, L., {Charmandaris}, V., {Georgakakis}, A., {et~al.} 2015, \aap, 576,
  A10

\bibitem[{{Cresci} {et~al.}(2015){Cresci}, {Mainieri}, {Brusa}, {Marconi},
  {Perna}, {Mannucci}, {Piconcelli}, {Maiolino}, {Feruglio}, {Fiore},
  {Bongiorno}, {Lanzuisi}, {Merloni}, {Schramm}, {Silverman}, \&
  {Civano}}]{Cresci2015a}
{Cresci}, G., {Mainieri}, V., {Brusa}, M., {et~al.} 2015, \apj, 799, 82

\bibitem[{{Crockett} {et~al.}(2012){Crockett}, {Shabala}, {Kaviraj},
  {Antonuccio-Delogu}, {Silk}, {Mutchler}, {O'Connell}, {Rejkuba}, {Whitmore},
  \& {Windhorst}}]{Crockett2012}
{Crockett}, R.~M., {Shabala}, S.~S., {Kaviraj}, S., {et~al.} 2012, \mnras, 421,
  1603

\bibitem[{{Croft} {et~al.}(2006){Croft}, {van Breugel}, {de Vries}, {Dopita},
  {Martin}, {Morganti}, {Neff}, {Oosterloo}, {Schiminovich}, {Stanford}, \&
  {van Gorkom}}]{Croft2006}
{Croft}, S., {van Breugel}, W., {de Vries}, W., {et~al.} 2006, \apj, 647, 1040

\bibitem[{{Croton} {et~al.}(2006){Croton}, {Springel}, {White}, {De Lucia},
  {Frenk}, {Gao}, {Jenkins}, {Kauffmann}, {Navarro}, \& {Yoshida}}]{Croton2006}
{Croton}, D.~J., {Springel}, V., {White}, S. D.~M., {et~al.} 2006, \mnras, 365,
  11

\bibitem[{{Cui} {et~al.}(2023){Cui}, {Hada}, {Kawashima}, {Kino}, {Lin},
  {Mizuno}, {Ro}, {Honma}, {Yi}, {Yu}, {Park}, {Jiang}, {Shen}, {Kravchenko},
  {Algaba}, {Cheng}, {Cho}, {Giovannini}, {Giroletti}, {Jung}, {Lu}, {Niinuma},
  {Oh}, {Ohsuga}, {Sawada-Satoh}, {Sohn}, {Takahashi}, {Takamura}, {Tazaki},
  {Trippe}, {Wajima}, {Akiyama}, {An}, {Asada}, {Buttaccio}, {Byun}, {Cui},
  {Hagiwara}, {Hirota}, {Hodgson}, {Kawaguchi}, {Kim}, {Lee}, {Lee}, {Lee},
  {Maccaferri}, {Melis}, {Melnikov}, {Migoni}, {Oh}, {Sugiyama}, {Wang},
  {Zhang}, {Chen}, {Hwang}, {Jung}, {Kim}, {Kim}, {Kobayashi}, {Li}, {Li},
  {Li}, {Liu}, {Liu}, {Liu}, {Oh}, {Oyama}, {Roh}, {Wang}, {Wang}, {Wang},
  {Xia}, {Yan}, {Yeom}, {Yonekura}, {Yuan}, {Zhang}, {Zhao}, \&
  {Zhong}}]{Cui2023}
{Cui}, Y., {Hada}, K., {Kawashima}, T., {et~al.} 2023, \nat, 621, 711

\bibitem[{{Dale} {et~al.}(2014){Dale}, {Helou}, {Magdis}, {Armus},
  {D{\'\i}az-Santos}, \& {Shi}}]{Dale2014}
{Dale}, D.~A., {Helou}, G., {Magdis}, G.~E., {et~al.} 2014, \apj, 784, 83

\bibitem[{{DeGraf} {et~al.}(2015){DeGraf}, {Di Matteo}, {Treu}, {Feng}, {Woo},
  \& {Park}}]{DeGraf2015}
{DeGraf}, C., {Di Matteo}, T., {Treu}, T., {et~al.} 2015, \mnras, 454, 913

\bibitem[{{Di Matteo} {et~al.}(2003{\natexlab{a}}){Di Matteo}, {Allen},
  {Fabian}, {Wilson}, \& {Young}}]{DiMatteo2003a}
{Di Matteo}, T., {Allen}, S.~W., {Fabian}, A.~C., {Wilson}, A.~S., \& {Young},
  A.~J. 2003{\natexlab{a}}, \apj, 582, 133

\bibitem[{{Di Matteo} {et~al.}(2008){Di Matteo}, {Colberg}, {Springel},
  {Hernquist}, \& {Sijacki}}]{DiMatteo2008}
{Di Matteo}, T., {Colberg}, J., {Springel}, V., {Hernquist}, L., \& {Sijacki},
  D. 2008, \apj, 676, 33

\bibitem[{{Di Matteo} {et~al.}(2003{\natexlab{b}}){Di Matteo}, {Croft},
  {Springel}, \& {Hernquist}}]{DiMatteo2003b}
{Di Matteo}, T., {Croft}, R. A.~C., {Springel}, V., \& {Hernquist}, L.
  2003{\natexlab{b}}, \apj, 593, 56

\bibitem[{{Di Matteo} {et~al.}(2004){Di Matteo}, {Croft}, {Springel}, \&
  {Hernquist}}]{DiMatteo2004}
{Di Matteo}, T., {Croft}, R. A.~C., {Springel}, V., \& {Hernquist}, L. 2004,
  \apj, 610, 80

\bibitem[{{Di Matteo} {et~al.}(2001){Di Matteo}, {Johnstone}, {Allen}, \&
  {Fabian}}]{DiMatteo2001}
{Di Matteo}, T., {Johnstone}, R.~M., {Allen}, S.~W., \& {Fabian}, A.~C. 2001,
  \apjl, 550, L19

\bibitem[{{Di Matteo} {et~al.}(2012){Di Matteo}, {Khandai}, {DeGraf}, {Feng},
  {Croft}, {Lopez}, \& {Springel}}]{DiMatteo2012}
{Di Matteo}, T., {Khandai}, N., {DeGraf}, C., {et~al.} 2012, \apjl, 745, L29

\bibitem[{{Di Matteo} {et~al.}(2000){Di Matteo}, {Quataert}, {Allen},
  {Narayan}, \& {Fabian}}]{DiMatteo2000}
{Di Matteo}, T., {Quataert}, E., {Allen}, S.~W., {Narayan}, R., \& {Fabian},
  A.~C. 2000, \mnras, 311, 507

\bibitem[{{Di Matteo} {et~al.}(2005){Di Matteo}, {Springel}, \&
  {Hernquist}}]{Di2005}
{Di Matteo}, T., {Springel}, V., \& {Hernquist}, L. 2005, \nat, 433, 604

\bibitem[{{Dotti} {et~al.}(2013){Dotti}, {Colpi}, {Pallini}, {Perego}, \&
  {Volonteri}}]{Dotti2013}
{Dotti}, M., {Colpi}, M., {Pallini}, S., {Perego}, A., \& {Volonteri}, M. 2013,
  \apj, 762, 68

\bibitem[{{Draghis} {et~al.}(2021){Draghis}, {Miller}, {Zoghbi}, {Kammoun},
  {Reynolds}, \& {Tomsick}}]{Draghis2021}
{Draghis}, P.~A., {Miller}, J.~M., {Zoghbi}, A., {et~al.} 2021, \apj, 920, 88

\bibitem[{{Draghis} {et~al.}(2023){Draghis}, {Miller}, {Zoghbi}, {Reynolds},
  {Costantini}, {Gallo}, \& {Tomsick}}]{Draghis2023}
{Draghis}, P.~A., {Miller}, J.~M., {Zoghbi}, A., {et~al.} 2023, \apj, 946, 19

\bibitem[{{Draine} \& {Li}(2007)}]{Draine2007}
{Draine}, B.~T. \& {Li}, A. 2007, \apj, 657, 810

\bibitem[{{Dubois} {et~al.}(2021){Dubois}, {Beckmann}, {Bournaud}, {Choi},
  {Devriendt}, {Jackson}, {Kaviraj}, {Kimm}, {Kraljic}, {Laigle}, {Martin},
  {Park}, {Peirani}, {Pichon}, {Volonteri}, \& {Yi}}]{Dubois2021}
{Dubois}, Y., {Beckmann}, R., {Bournaud}, F., {et~al.} 2021, \aap, 651, A109

\bibitem[{{Dubois} {et~al.}(2013){Dubois}, {Gavazzi}, {Peirani}, \&
  {Silk}}]{Dubois2013}
{Dubois}, Y., {Gavazzi}, R., {Peirani}, S., \& {Silk}, J. 2013, \mnras, 433,
  3297

\bibitem[{{Dubois} {et~al.}(2016){Dubois}, {Peirani}, {Pichon}, {Devriendt},
  {Gavazzi}, {Welker}, \& {Volonteri}}]{Dubois2016}
{Dubois}, Y., {Peirani}, S., {Pichon}, C., {et~al.} 2016, \mnras, 463, 3948

\bibitem[{{Dubois} {et~al.}(2014){Dubois}, {Volonteri}, \& {Silk}}]{Dubois2014}
{Dubois}, Y., {Volonteri}, M., \& {Silk}, J. 2014, \mnras, 440, 1590

\bibitem[{{Elbaz} {et~al.}(2009){Elbaz}, {Jahnke}, {Pantin}, {Le Borgne}, \&
  {Letawe}}]{Elbaz2009}
{Elbaz}, D., {Jahnke}, K., {Pantin}, E., {Le Borgne}, D., \& {Letawe}, G. 2009,
  \aap, 507, 1359

\bibitem[{{Ellison} {et~al.}(2013){Ellison}, {Mendel}, {Patton}, \&
  {Scudder}}]{Ellison2013}
{Ellison}, S.~L., {Mendel}, J.~T., {Patton}, D.~R., \& {Scudder}, J.~M. 2013,
  \mnras, 435, 3627

\bibitem[{{Emonts} {et~al.}(2005){Emonts}, {Morganti}, {Tadhunter},
  {Oosterloo}, {Holt}, \& {van der Hulst}}]{Emonts2005}
{Emonts}, B.~H.~C., {Morganti}, R., {Tadhunter}, C.~N., {et~al.} 2005, \mnras,
  362, 931

\bibitem[{{Fabian}(2012)}]{Fabian2012}
{Fabian}, A.~C. 2012, \araa, 50, 455

\bibitem[{{Fanidakis} {et~al.}(2011){Fanidakis}, {Baugh}, {Benson}, {Bower},
  {Cole}, {Done}, \& {Frenk}}]{Fanidakis2011}
{Fanidakis}, N., {Baugh}, C.~M., {Benson}, A.~J., {et~al.} 2011, \mnras, 410,
  53

\bibitem[{{Feain} {et~al.}(2007){Feain}, {Papadopoulos}, {Ekers}, \&
  {Middelberg}}]{Feain2007}
{Feain}, I.~J., {Papadopoulos}, P.~P., {Ekers}, R.~D., \& {Middelberg}, E.
  2007, \apj, 662, 872

\bibitem[{{Ferrarese} {et~al.}(2006){Ferrarese}, {C{\^o}t{\'e}}, {Dalla
  Bont{\`a}}, {Peng}, {Merritt}, {Jord{\'a}n}, {Blakeslee}, {Ha{\c{s}}egan},
  {Mei}, {Piatek}, {Tonry}, \& {West}}]{Ferrarese2006}
{Ferrarese}, L., {C{\^o}t{\'e}}, P., {Dalla Bont{\`a}}, E., {et~al.} 2006,
  \apjl, 644, L21

\bibitem[{{Ferrarese} \& {Merritt}(2000)}]{Ferrarese2000}
{Ferrarese}, L. \& {Merritt}, D. 2000, \apjl, 539, L9

\bibitem[{{Ferrarese} {et~al.}(2001){Ferrarese}, {Pogge}, {Peterson},
  {Merritt}, {Wandel}, \& {Joseph}}]{Ferrarese2001}
{Ferrarese}, L., {Pogge}, R.~W., {Peterson}, B.~M., {et~al.} 2001, \apjl, 555,
  L79

\bibitem[{{Fritz} {et~al.}(2006){Fritz}, {Franceschini}, \&
  {Hatziminaoglou}}]{Fritz2006}
{Fritz}, J., {Franceschini}, A., \& {Hatziminaoglou}, E. 2006, \mnras, 366, 767

\bibitem[{{Gaibler} {et~al.}(2012){Gaibler}, {Khochfar}, {Krause}, \&
  {Silk}}]{Gaibler2012}
{Gaibler}, V., {Khochfar}, S., {Krause}, M., \& {Silk}, J. 2012, \mnras, 425,
  438

\bibitem[{{Gallo} {et~al.}(2011){Gallo}, {Miniutti}, {Miller}, {Brenneman},
  {Fabian}, {Guainazzi}, \& {Reynolds}}]{Gallo2011}
{Gallo}, L.~C., {Miniutti}, G., {Miller}, J.~M., {et~al.} 2011, \mnras, 411,
  607

\bibitem[{{Garofalo} {et~al.}(2010){Garofalo}, {Evans}, \&
  {Sambruna}}]{Garofalo2010}
{Garofalo}, D., {Evans}, D.~A., \& {Sambruna}, R.~M. 2010, \mnras, 406, 975

\bibitem[{{Gebhardt} {et~al.}(2000){Gebhardt}, {Bender}, {Bower}, {Dressler},
  {Faber}, {Filippenko}, {Green}, {Grillmair}, {Ho}, {Kormendy}, {Lauer},
  {Magorrian}, {Pinkney}, {Richstone}, \& {Tremaine}}]{Gebhardt2000}
{Gebhardt}, K., {Bender}, R., {Bower}, G., {et~al.} 2000, \apjl, 539, L13

\bibitem[{{Ghisellini}(2006)}]{Ghisellini2006}
{Ghisellini}, G. 2006, in VI Microquasar Workshop: Microquasars and Beyond,
  27.1

\bibitem[{{Greene} {et~al.}(2010){Greene}, {Peng}, {Kim}, {Kuo}, {Braatz},
  {Impellizzeri}, {Condon}, {Lo}, {Henkel}, \& {Reid}}]{Greene2010}
{Greene}, J.~E., {Peng}, C.~Y., {Kim}, M., {et~al.} 2010, \apj, 721, 26

\bibitem[{{Griffin} {et~al.}(2019){Griffin}, {Lacey}, {Gonzalez-Perez},
  {Lagos}, {Baugh}, \& {Fanidakis}}]{Griffin2019}
{Griffin}, A.~J., {Lacey}, C.~G., {Gonzalez-Perez}, V., {et~al.} 2019, \mnras,
  487, 198

\bibitem[{{G{\"u}ltekin} {et~al.}(2009){G{\"u}ltekin}, {Richstone}, {Gebhardt},
  {Lauer}, {Tremaine}, {Aller}, {Bender}, {Dressler}, {Faber}, {Filippenko},
  {Green}, {Ho}, {Kormendy}, {Magorrian}, {Pinkney}, \&
  {Siopis}}]{Gultekin2009}
{G{\"u}ltekin}, K., {Richstone}, D.~O., {Gebhardt}, K., {et~al.} 2009, \apj,
  698, 198

\bibitem[{{H{\"a}ring} \& {Rix}(2004)}]{Haring2004}
{H{\"a}ring}, N. \& {Rix}, H.-W. 2004, \apjl, 604, L89

\bibitem[{{Heckman} \& {Best}(2014)}]{Heckman2014}
{Heckman}, T.~M. \& {Best}, P.~N. 2014, \araa, 52, 589

\bibitem[{{Ho}(2008)}]{Ho2008}
{Ho}, L.~C. 2008, \araa, 46, 475

\bibitem[{{Hopkins} {et~al.}(2006){Hopkins}, {Hernquist}, {Cox}, {Di Matteo},
  {Robertson}, \& {Springel}}]{Hopkins2006}
{Hopkins}, P.~F., {Hernquist}, L., {Cox}, T.~J., {et~al.} 2006, \apjs, 163, 1

\bibitem[{{Hopkins} {et~al.}(2016){Hopkins}, {Torrey}, {Faucher-Gigu{\`e}re},
  {Quataert}, \& {Murray}}]{Hopkins2016}
{Hopkins}, P.~F., {Torrey}, P., {Faucher-Gigu{\`e}re}, C.-A., {Quataert}, E.,
  \& {Murray}, N. 2016, \mnras, 458, 816

\bibitem[{{Imanishi} {et~al.}(2011){Imanishi}, {Ichikawa}, {Takeuchi},
  {Kawakatu}, {Oi}, \& {Imase}}]{Imanishi2011}
{Imanishi}, M., {Ichikawa}, K., {Takeuchi}, T., {et~al.} 2011, \pasj, 63, 447

\bibitem[{{Ishibashi} \& {Fabian}(2012)}]{Ishibashi2012}
{Ishibashi}, W. \& {Fabian}, A.~C. 2012, \mnras, 427, 2998

\bibitem[{{Ishibashi} {et~al.}(2019){Ishibashi}, {Fabian}, \&
  {Reynolds}}]{Ishibashi2019}
{Ishibashi}, W., {Fabian}, A.~C., \& {Reynolds}, C.~S. 2019, \mnras, 486, 2210

\bibitem[{{Izquierdo-Villalba} {et~al.}(2020){Izquierdo-Villalba}, {Bonoli},
  {Dotti}, {Sesana}, {Rosas-Guevara}, \& {Spinoso}}]{Izquierdo2020}
{Izquierdo-Villalba}, D., {Bonoli}, S., {Dotti}, M., {et~al.} 2020, \mnras,
  495, 4681

\bibitem[{{Jana} {et~al.}(2021){Jana}, {Naik}, {Chatterjee}, \&
  {Jaisawal}}]{Jana2021}
{Jana}, A., {Naik}, S., {Chatterjee}, D., \& {Jaisawal}, G.~K. 2021, \mnras,
  507, 4779

\bibitem[{{Jia} {et~al.}(2022){Jia}, {Zhao}, {Gou}, {Garc{\'\i}a}, {Liao},
  {Feng}, {Li}, {Wang}, {Li}, \& {Wu}}]{Jia2022}
{Jia}, N., {Zhao}, X., {Gou}, L., {et~al.} 2022, \mnras, 511, 3125

\bibitem[{{Jiang} {et~al.}(2019){Jiang}, {Walton}, {Fabian}, \&
  {Parker}}]{Jiang2019}
{Jiang}, J., {Walton}, D.~J., {Fabian}, A.~C., \& {Parker}, M.~L. 2019, \mnras,
  483, 2958

\bibitem[{{Kauffmann} \& {Haehnelt}(2000)}]{Kauffmann2000}
{Kauffmann}, G. \& {Haehnelt}, M. 2000, \mnras, 311, 576

\bibitem[{{Keck} {et~al.}(2015){Keck}, {Brenneman}, {Ballantyne}, {Bauer},
  {Boggs}, {Christensen}, {Craig}, {Dauser}, {Elvis}, {Fabian}, {Fuerst},
  {Garc{\'\i}a}, {Grefenstette}, {Hailey}, {Harrison}, {Madejski}, {Marinucci},
  {Matt}, {Reynolds}, {Stern}, {Walton}, \& {Zoghbi}}]{Keck2015}
{Keck}, M.~L., {Brenneman}, L.~W., {Ballantyne}, D.~R., {et~al.} 2015, \apj,
  806, 149

\bibitem[{{Kelly}(2007)}]{Kelly2007}
{Kelly}, B.~C. 2007, \apj, 665, 1489

\bibitem[{{Kennicutt}(1998)}]{Kennicut1998}
{Kennicutt}, Robert~C., J. 1998, \araa, 36, 189

\bibitem[{{Kim} {et~al.}(2009){Kim}, {Wise}, \& {Abel}}]{Kim2009}
{Kim}, J.-h., {Wise}, J.~H., \& {Abel}, T. 2009, \apjl, 694, L123

\bibitem[{{Knapen} {et~al.}(2015){Knapen}, {Cisternas}, \&
  {Querejeta}}]{Knapen2015}
{Knapen}, J.~H., {Cisternas}, M., \& {Querejeta}, M. 2015, \mnras, 454, 1742

\bibitem[{{Koide} {et~al.}(2002){Koide}, {Shibata}, {Kudoh}, \&
  {Meier}}]{Koide2002}
{Koide}, S., {Shibata}, K., {Kudoh}, T., \& {Meier}, D.~L. 2002, Science, 295,
  1688

\bibitem[{{Kormendy} \& {Gebhardt}(2001)}]{Kormendy2001}
{Kormendy}, J. \& {Gebhardt}, K. 2001, in American Institute of Physics
  Conference Series, Vol. 586, 20th Texas Symposium on relativistic
  astrophysics, ed. J.~C. {Wheeler} \& H.~{Martel} (AIP), 363--381

\bibitem[{{Kormendy} \& {Ho}(2013)}]{Kormendy2013}
{Kormendy}, J. \& {Ho}, L.~C. 2013, \araa, 51, 511

\bibitem[{{Kormendy} \& {Richstone}(1995)}]{Kormendy1995}
{Kormendy}, J. \& {Richstone}, D. 1995, \araa, 33, 581

\bibitem[{{Koss} {et~al.}(2022){Koss}, {Ricci}, {Trakhtenbrot}, {Oh}, {den
  Brok}, {Mej{\'\i}a-Restrepo}, {Stern}, {Privon}, {Treister}, {Powell},
  {Mushotzky}, {Bauer}, {Ananna}, {Balokovi{\'c}}, {B{\"a}r}, {Becker},
  {Bessiere}, {Burtscher}, {Caglar}, {Congiu}, {Evans}, {Harrison}, {Heida},
  {Ichikawa}, {Kamraj}, {Lamperti}, {Pacucci}, {Ricci}, {Riffel}, {Rojas},
  {Schawinski}, {Temple}, {Urry}, {Veilleux}, \& {Williams}}]{Koss2022}
{Koss}, M.~J., {Ricci}, C., {Trakhtenbrot}, B., {et~al.} 2022, \apjs, 261, 2

\bibitem[{{Kroupa}(2001)}]{Kroupa2001}
{Kroupa}, P. 2001, \mnras, 322, 231

\bibitem[{{Lammers} {et~al.}(2023){Lammers}, {Iyer}, {Ibarra-Medel},
  {Pacifici}, {S{\'a}nchez}, {Tacchella}, \& {Woo}}]{Lammers2023}
{Lammers}, C., {Iyer}, K.~G., {Ibarra-Medel}, H., {et~al.} 2023, \apj, 953, 26

\bibitem[{{Lin} {et~al.}(2008){Lin}, {Patton}, {Koo}, {Casteels}, {Conselice},
  {Faber}, {Lotz}, {Willmer}, {Hsieh}, {Chiueh}, {Newman}, {Novak}, {Weiner},
  \& {Cooper}}]{Lin2008}
{Lin}, L., {Patton}, D.~R., {Koo}, D.~C., {et~al.} 2008, \apj, 681, 232

\bibitem[{{Liu} {et~al.}(2013){Liu}, {Gan}, \& {Xie}}]{Liu2013}
{Liu}, C., {Gan}, Z.-M., \& {Xie}, F.-G. 2013, Research in Astronomy and
  Astrophysics, 13, 899

\bibitem[{{Livio} {et~al.}(1999){Livio}, {Ogilvie}, \& {Pringle}}]{Livio1999}
{Livio}, M., {Ogilvie}, G.~I., \& {Pringle}, J.~E. 1999, \apj, 512, 100

\bibitem[{{Lohfink} {et~al.}(2013){Lohfink}, {Reynolds}, {Jorstad}, {Marscher},
  {Miller}, {Aller}, {Aller}, {Brenneman}, {Fabian}, {Miller}, {Mushotzky},
  {Nowak}, \& {Tombesi}}]{Lohfink2013}
{Lohfink}, A.~M., {Reynolds}, C.~S., {Jorstad}, S.~G., {et~al.} 2013, \apj,
  772, 83

\bibitem[{{Lu} {et~al.}(2019){Lu}, {Huang}, {Zhang}, {Wang}, {Du}, {Hu},
  {Xiao}, {Li}, {Bai}, {Bian}, {Yuan}, {Ho}, {Wang}, \& {SEAMBH
  Collaboration}}]{Lu2019}
{Lu}, K.-X., {Huang}, Y.-K., {Zhang}, Z.-X., {et~al.} 2019, \apj, 877, 23

\bibitem[{{Lynden-Bell}(1969)}]{LyndenBell1969}
{Lynden-Bell}, D. 1969, \nat, 223, 690

\bibitem[{{MacDonald} \& {Thorne}(1982)}]{Macdonald82}
{MacDonald}, D. \& {Thorne}, K.~S. 1982, \mnras, 198, 345

\bibitem[{{Magnier} {et~al.}(2020){Magnier}, {Chambers}, {Flewelling},
  {Hoblitt}, {Huber}, {Price}, {Sweeney}, {Waters}, {Denneau}, {Draper},
  {Hodapp}, {Jedicke}, {Kaiser}, {Kudritzki}, {Metcalfe}, {Stubbs}, \&
  {Wainscoat}}]{Magnier2020}
{Magnier}, E.~A., {Chambers}, K.~C., {Flewelling}, H.~A., {et~al.} 2020, \apjs,
  251, 3

\bibitem[{{Magorrian} {et~al.}(1998){Magorrian}, {Tremaine}, {Richstone},
  {Bender}, {Bower}, {Dressler}, {Faber}, {Gebhardt}, {Green}, {Grillmair},
  {Kormendy}, \& {Lauer}}]{Magorrian1998}
{Magorrian}, J., {Tremaine}, S., {Richstone}, D., {et~al.} 1998, \aj, 115, 2285

\bibitem[{{Mahony} {et~al.}(2016){Mahony}, {Oonk}, {Morganti}, {Tadhunter},
  {Bessiere}, {Short}, {Emonts}, \& {Oosterloo}}]{Mahony2016}
{Mahony}, E.~K., {Oonk}, J.~B.~R., {Morganti}, R., {et~al.} 2016, \mnras, 455,
  2453

\bibitem[{{Mallick} {et~al.}(2022){Mallick}, {Fabian}, {Garc{\'\i}a},
  {Tomsick}, {Parker}, {Dauser}, {Wilkins}, {De Marco}, {Steiner}, {Connors},
  {Mastroserio}, {Markowitz}, {Pinto}, {Alston}, {Lohfink}, \&
  {Gandhi}}]{Mallick2022}
{Mallick}, L., {Fabian}, A.~C., {Garc{\'\i}a}, J.~A., {et~al.} 2022, \mnras,
  513, 4361

\bibitem[{{Marconi} \& {Hunt}(2003)}]{Marconi2003}
{Marconi}, A. \& {Hunt}, L.~K. 2003, \apjl, 589, L21

\bibitem[{{McKinney}(2005)}]{McKinney2005}
{McKinney}, J.~C. 2005, \apjl, 630, L5

\bibitem[{{McLure} \& {Dunlop}(2001)}]{McLure2001}
{McLure}, R.~J. \& {Dunlop}, J.~S. 2001, \mnras, 327, 199

\bibitem[{{McLure} \& {Dunlop}(2002)}]{McLure2002}
{McLure}, R.~J. \& {Dunlop}, J.~S. 2002, \mnras, 331, 795

\bibitem[{{Meier}(2002)}]{Meier2002}
{Meier}, D.~L. 2002, \nar, 46, 247

\bibitem[{{Molina} {et~al.}(2023){Molina}, {Shangguan}, {Wang}, {Ho}, {Bauer},
  \& {Treister}}]{Molina2023}
{Molina}, J., {Shangguan}, J., {Wang}, R., {et~al.} 2023, \apj, 950, 60

\bibitem[{{Morganti} {et~al.}(2013{\natexlab{a}}){Morganti}, {Fogasy},
  {Paragi}, {Oosterloo}, \& {Orienti}}]{Morganti2013a}
{Morganti}, R., {Fogasy}, J., {Paragi}, Z., {Oosterloo}, T., \& {Orienti}, M.
  2013{\natexlab{a}}, Science, 341, 1082

\bibitem[{{Morganti} {et~al.}(2013{\natexlab{b}}){Morganti}, {Frieswijk},
  {Oonk}, {Oosterloo}, \& {Tadhunter}}]{Morganti2013b}
{Morganti}, R., {Frieswijk}, W., {Oonk}, R.~J.~B., {Oosterloo}, T., \&
  {Tadhunter}, C. 2013{\natexlab{b}}, \aap, 552, L4

\bibitem[{{Morganti} {et~al.}(2007){Morganti}, {Holt}, {Saripalli},
  {Oosterloo}, \& {Tadhunter}}]{Morganti2007}
{Morganti}, R., {Holt}, J., {Saripalli}, L., {Oosterloo}, T.~A., \&
  {Tadhunter}, C.~N. 2007, \aap, 476, 735

\bibitem[{{Narayan} \& {McClintock}(2012)}]{Narayan2012}
{Narayan}, R. \& {McClintock}, J.~E. 2012, \mnras, 419, L69

\bibitem[{{Niko{\l}ajuk} {et~al.}(2009){Niko{\l}ajuk}, {Czerny}, \&
  {Gurynowicz}}]{Nikolajuk2009}
{Niko{\l}ajuk}, M., {Czerny}, B., \& {Gurynowicz}, P. 2009, \mnras, 394, 2141

\bibitem[{{Noll} {et~al.}(2009){Noll}, {Burgarella}, {Giovannoli}, {Buat},
  {Marcillac}, \& {Mu{\~n}oz-Mateos}}]{Noll2009}
{Noll}, S., {Burgarella}, D., {Giovannoli}, E., {et~al.} 2009, \aap, 507, 1793

\bibitem[{{Pacucci} \& {Loeb}(2020)}]{Pacucci2020}
{Pacucci}, F. \& {Loeb}, A. 2020, \apj, 895, 95

\bibitem[{{Pacucci} {et~al.}(2023){Pacucci}, {Nguyen}, {Carniani}, {Maiolino},
  \& {Fan}}]{Pacucci2023}
{Pacucci}, F., {Nguyen}, B., {Carniani}, S., {Maiolino}, R., \& {Fan}, X. 2023,
  \apjl, 957, L3

\bibitem[{{Parker} {et~al.}(2018){Parker}, {Matzeu}, {Guainazzi},
  {Kalfountzou}, {Miniutti}, {Santos-Lle{\'o}}, \& {Schartel}}]{Parker2018}
{Parker}, M.~L., {Matzeu}, G.~A., {Guainazzi}, M., {et~al.} 2018, \mnras, 480,
  2365

\bibitem[{{Peirani} {et~al.}(2024){Peirani}, {Suto}, {Beckmann}, {Volonteri},
  {Lin}, {Dubois}, {Yi}, {Pichon}, {Kraljic}, {Park}, {Devriendt}, {Han}, \&
  {Chen}}]{Peirani2024}
{Peirani}, S., {Suto}, Y., {Beckmann}, R.~S., {et~al.} 2024, \aap, 686, A233

\bibitem[{{Perez} {et~al.}(2011){Perez}, {Michel-Dansac}, \&
  {Tissera}}]{Perez2011}
{Perez}, J., {Michel-Dansac}, L., \& {Tissera}, P.~B. 2011, \mnras, 417, 580

\bibitem[{{Peterson} {et~al.}(2004){Peterson}, {Ferrarese}, {Gilbert}, {Kaspi},
  {Malkan}, {Maoz}, {Merritt}, {Netzer}, {Onken}, {Pogge}, {Vestergaard}, \&
  {Wandel}}]{Peterson2004}
{Peterson}, B.~M., {Ferrarese}, L., {Gilbert}, K.~M., {et~al.} 2004, \apj, 613,
  682

\bibitem[{{Randall} {et~al.}(2008){Randall}, {Markevitch}, {Clowe}, {Gonzalez},
  \& {Brada{\v{c}}}}]{Randall2008}
{Randall}, S.~W., {Markevitch}, M., {Clowe}, D., {Gonzalez}, A.~H., \&
  {Brada{\v{c}}}, M. 2008, \apj, 679, 1173

\bibitem[{{Reines} \& {Volonteri}(2015)}]{Reines2015}
{Reines}, A.~E. \& {Volonteri}, M. 2015, \apj, 813, 82

\bibitem[{{Reis} {et~al.}(2014){Reis}, {Reynolds}, {Miller}, \&
  {Walton}}]{Reis2014}
{Reis}, R.~C., {Reynolds}, M.~T., {Miller}, J.~M., \& {Walton}, D.~J. 2014,
  \nat, 507, 207

\bibitem[{{Reynolds}(2021)}]{Reynolds2021}
{Reynolds}, C.~S. 2021, \araa, 59, 117

\bibitem[{{Risaliti} {et~al.}(2013){Risaliti}, {Harrison}, {Madsen}, {Walton},
  {Boggs}, {Christensen}, {Craig}, {Grefenstette}, {Hailey}, {Nardini},
  {Stern}, \& {Zhang}}]{Risaliti2013}
{Risaliti}, G., {Harrison}, F.~A., {Madsen}, K.~K., {et~al.} 2013, \nat, 494,
  449

\bibitem[{{Rodr{\'\i}guez Zaur{\'\i}n} {et~al.}(2007){Rodr{\'\i}guez
  Zaur{\'\i}n}, {Holt}, {Tadhunter}, \& {Gonz{\'a}lez Delgado}}]{Rod2007}
{Rodr{\'\i}guez Zaur{\'\i}n}, J., {Holt}, J., {Tadhunter}, C.~N., \&
  {Gonz{\'a}lez Delgado}, R.~M. 2007, \mnras, 375, 1133

\bibitem[{{Rupke} {et~al.}(2010){Rupke}, {Kewley}, \& {Barnes}}]{Rupke2010}
{Rupke}, D. S.~N., {Kewley}, L.~J., \& {Barnes}, J.~E. 2010, \apjl, 710, L156

\bibitem[{{Saitoh} {et~al.}(2009){Saitoh}, {Daisaka}, {Kokubo}, {Makino},
  {Okamoto}, {Tomisaka}, {Wada}, \& {Yoshida}}]{Saitoh2009}
{Saitoh}, T.~R., {Daisaka}, H., {Kokubo}, E., {et~al.} 2009, \pasj, 61, 481

\bibitem[{{Sala} {et~al.}(2024){Sala}, {Valentini}, {Biffi}, \&
  {Dolag}}]{Sala2024}
{Sala}, L., {Valentini}, M., {Biffi}, V., \& {Dolag}, K. 2024, \aap, 685, A92

\bibitem[{{Santini} {et~al.}(2012){Santini}, {Rosario}, {Shao}, {Lutz},
  {Maiolino}, {Alexander}, {Altieri}, {Andreani}, {Aussel}, {Bauer}, {Berta},
  {Bongiovanni}, {Brandt}, {Brusa}, {Cepa}, {Cimatti}, {Daddi}, {Elbaz},
  {Fontana}, {F{\"o}rster Schreiber}, {Genzel}, {Grazian}, {Le Floc'h},
  {Magnelli}, {Mainieri}, {Nordon}, {P{\'e}rez Garcia}, {Poglitsch}, {Popesso},
  {Pozzi}, {Riguccini}, {Rodighiero}, {Salvato}, {Sanchez-Portal}, {Sturm},
  {Tacconi}, {Valtchanov}, \& {Wuyts}}]{Santini2012}
{Santini}, P., {Rosario}, D.~J., {Shao}, L., {et~al.} 2012, \aap, 540, A109

\bibitem[{{Schaye} {et~al.}(2015){Schaye}, {Crain}, {Bower}, {Furlong},
  {Schaller}, {Theuns}, {Dalla Vecchia}, {Frenk}, {McCarthy}, {Helly},
  {Jenkins}, {Rosas-Guevara}, {White}, {Baes}, {Booth}, {Camps}, {Navarro},
  {Qu}, {Rahmati}, {Sawala}, {Thomas}, \& {Trayford}}]{Schaye2015}
{Schaye}, J., {Crain}, R.~A., {Bower}, R.~G., {et~al.} 2015, \mnras, 446, 521

\bibitem[{{Schutte} \& {Reines}(2022)}]{Schutte2022}
{Schutte}, Z. \& {Reines}, A.~E. 2022, \nat, 601, 329

\bibitem[{{Sesana} {et~al.}(2014){Sesana}, {Barausse}, {Dotti}, \&
  {Rossi}}]{Sesana2014}
{Sesana}, A., {Barausse}, E., {Dotti}, M., \& {Rossi}, E.~M. 2014, \apj, 794,
  104

\bibitem[{{Shakura} \& {Sunyaev}(1976)}]{Shakura1976}
{Shakura}, N.~I. \& {Sunyaev}, R.~A. 1976, \mnras, 175, 613

\bibitem[{{Sijacki} {et~al.}(2007){Sijacki}, {Springel}, {Di Matteo}, \&
  {Hernquist}}]{Sijacki2007}
{Sijacki}, D., {Springel}, V., {Di Matteo}, T., \& {Hernquist}, L. 2007,
  \mnras, 380, 877

\bibitem[{{Silk}(2013)}]{Silk2013}
{Silk}, J. 2013, \apj, 772, 112

\bibitem[{{Silk} \& {Norman}(2009)}]{Silk2009}
{Silk}, J. \& {Norman}, C. 2009, \apj, 700, 262

\bibitem[{{Silva} {et~al.}(2018){Silva}, {Marchesini}, {Silverman}, {Skelton},
  {Iono}, {Martis}, {Marsan}, {Tadaki}, {Brammer}, \& {kartaltepe}}]{Silva2018}
{Silva}, A., {Marchesini}, D., {Silverman}, J.~D., {et~al.} 2018, \apj, 868, 46

\bibitem[{{Skrutskie} {et~al.}(2006){Skrutskie}, {Cutri}, {Stiening},
  {Weinberg}, {Schneider}, {Carpenter}, {Beichman}, {Capps}, {Chester},
  {Elias}, {Huchra}, {Liebert}, {Lonsdale}, {Monet}, {Price}, {Seitzer},
  {Jarrett}, {Kirkpatrick}, {Gizis}, {Howard}, {Evans}, {Fowler}, {Fullmer},
  {Hurt}, {Light}, {Kopan}, {Marsh}, {McCallon}, {Tam}, {Van Dyk}, \&
  {Wheelock}}]{Skrutskie2006}
{Skrutskie}, M.~F., {Cutri}, R.~M., {Stiening}, R., {et~al.} 2006, \aj, 131,
  1163

\bibitem[{{Sluse} {et~al.}(2012){Sluse}, {Hutsem{\'e}kers}, {Courbin},
  {Meylan}, \& {Wambsganss}}]{Sluse2012}
{Sluse}, D., {Hutsem{\'e}kers}, D., {Courbin}, F., {Meylan}, G., \&
  {Wambsganss}, J. 2012, \aap, 544, A62

\bibitem[{{Soares} \& {Nemmen}(2020)}]{Soares2020}
{Soares}, G. \& {Nemmen}, R. 2020, \mnras, 495, 981

\bibitem[{{Soltan}(1982)}]{Soltan1982}
{Soltan}, A. 1982, \mnras, 200, 115

\bibitem[{{Springel} {et~al.}(2005){Springel}, {Di Matteo}, \&
  {Hernquist}}]{Springel2005}
{Springel}, V., {Di Matteo}, T., \& {Hernquist}, L. 2005, \mnras, 361, 776

\bibitem[{{Steiner} {et~al.}(2013){Steiner}, {McClintock}, \&
  {Narayan}}]{Steiner2013}
{Steiner}, J.~F., {McClintock}, J.~E., \& {Narayan}, R. 2013, \apj, 762, 104

\bibitem[{{Suh} {et~al.}(2019){Suh}, {Civano}, {Hasinger}, {Lusso}, {Marchesi},
  {Schulze}, {Onodera}, {Rosario}, \& {Sanders}}]{Suh2019}
{Suh}, H., {Civano}, F., {Hasinger}, G., {et~al.} 2019, \apj, 872, 168

\bibitem[{{Sun} {et~al.}(2018){Sun}, {Guainazzi}, {Ni}, {Wang}, {Qian}, {Shi},
  {Wang}, \& {Bambi}}]{Sun2018}
{Sun}, S., {Guainazzi}, M., {Ni}, Q., {et~al.} 2018, \mnras, 478, 1900

\bibitem[{{Tadhunter} {et~al.}(2014){Tadhunter}, {Morganti}, {Rose}, {Oonk}, \&
  {Oosterloo}}]{Tadhunter2014}
{Tadhunter}, C., {Morganti}, R., {Rose}, M., {Oonk}, J.~B.~R., \& {Oosterloo},
  T. 2014, \nat, 511, 440

\bibitem[{{Tanabe} \& {Nagataki}(2008)}]{Tanabe2008}
{Tanabe}, K. \& {Nagataki}, S. 2008, \prd, 78, 024004

\bibitem[{{Tchekhovskoy} {et~al.}(2012){Tchekhovskoy}, {McKinney}, \&
  {Narayan}}]{Tchekhovskoy12}
{Tchekhovskoy}, A., {McKinney}, J.~C., \& {Narayan}, R. 2012, in Journal of
  Physics Conference Series, Vol. 372, Journal of Physics Conference Series
  (IOP), 012040

\bibitem[{{Thorne} {et~al.}(1986){Thorne}, {Price}, \&
  {MacDonald}}]{Thorne1986}
{Thorne}, K.~S., {Price}, R.~H., \& {MacDonald}, D.~A. 1986, {Black holes: The
  membrane paradigm}

\bibitem[{{Tremaine} {et~al.}(2002){Tremaine}, {Gebhardt}, {Bender}, {Bower},
  {Dressler}, {Faber}, {Filippenko}, {Green}, {Grillmair}, {Ho}, {Kormendy},
  {Lauer}, {Magorrian}, {Pinkney}, \& {Richstone}}]{Tremaine2002}
{Tremaine}, S., {Gebhardt}, K., {Bender}, R., {et~al.} 2002, \apj, 574, 740

\bibitem[{{{\"U}nal} \& {Loeb}(2020)}]{Unal2020}
{{\"U}nal}, C. \& {Loeb}, A. 2020, \mnras, 495, 278

\bibitem[{{Vasudevan} {et~al.}(2016){Vasudevan}, {Fabian}, {Reynolds}, {Aird},
  {Dauser}, \& {Gallo}}]{Vasudevan2016}
{Vasudevan}, R.~V., {Fabian}, A.~C., {Reynolds}, C.~S., {et~al.} 2016, \mnras,
  458, 2012

\bibitem[{{Venturi} {et~al.}(2023){Venturi}, {Treister}, {Finlez}, {D'Ago},
  {Bauer}, {Harrison}, {Ramos Almeida}, {Revalski}, {Ricci}, {Sartori},
  {Girdhar}, {Keel}, \& {Tub{\'\i}n}}]{Venturi2023}
{Venturi}, G., {Treister}, E., {Finlez}, C., {et~al.} 2023, \aap, 678, A127

\bibitem[{{Volonteri} {et~al.}(2005){Volonteri}, {Madau}, {Quataert}, \&
  {Rees}}]{Volonteri2005}
{Volonteri}, M., {Madau}, P., {Quataert}, E., \& {Rees}, M.~J. 2005, \apj, 620,
  69

\bibitem[{{Volonteri} {et~al.}(2013){Volonteri}, {Sikora}, {Lasota}, \&
  {Merloni}}]{Volonteri2013}
{Volonteri}, M., {Sikora}, M., {Lasota}, J.~P., \& {Merloni}, A. 2013, \apj,
  775, 94

\bibitem[{{Walton} {et~al.}(2020){Walton}, {Alston}, {Kosec}, {Fabian},
  {Gallo}, {Garcia}, {Miller}, {Nardini}, {Reynolds}, {Ricci}, {Stern},
  {Dauser}, {Harrison}, \& {Reynolds}}]{Walton2020}
{Walton}, D.~J., {Alston}, W.~N., {Kosec}, P., {et~al.} 2020, \mnras, 499, 1480

\bibitem[{{Walton} {et~al.}(2013){Walton}, {Nardini}, {Fabian}, {Gallo}, \&
  {Reis}}]{Walton2013}
{Walton}, D.~J., {Nardini}, E., {Fabian}, A.~C., {Gallo}, L.~C., \& {Reis},
  R.~C. 2013, \mnras, 428, 2901

\bibitem[{{Walton} {et~al.}(2019){Walton}, {Nardini}, {Gallo}, {Reynolds},
  {Ricci}, {Dauser}, {Fabian}, {Garc{\'\i}a}, {Harrison}, {Risaliti}, \&
  {Stern}}]{Walton2019}
{Walton}, D.~J., {Nardini}, E., {Gallo}, L.~C., {et~al.} 2019, \mnras, 484,
  2544

\bibitem[{{Wang} {et~al.}(2018){Wang}, {M{\'e}ndez}, {Altamirano}, {Court},
  {Beri}, \& {Cheng}}]{Wang2018}
{Wang}, Y., {M{\'e}ndez}, M., {Altamirano}, D., {et~al.} 2018, \mnras, 478,
  4837

\bibitem[{{Weinberger} {et~al.}(2017){Weinberger}, {Springel}, {Hernquist},
  {Pillepich}, {Marinacci}, {Pakmor}, {Nelson}, {Genel}, {Vogelsberger},
  {Naiman}, \& {Torrey}}]{Weinberger2017}
{Weinberger}, R., {Springel}, V., {Hernquist}, L., {et~al.} 2017, \mnras, 465,
  3291

\bibitem[{{Wolter} {et~al.}(2023){Wolter}, {Berg}, \& {Chisholm}}]{Wolter2023}
{Wolter}, I.~E., {Berg}, M.~A., \& {Chisholm}, J. 2023, Research Notes of the
  American Astronomical Society, 7, 232

\bibitem[{{Wright} {et~al.}(2010){Wright}, {Eisenhardt}, {Mainzer}, {Ressler},
  {Cutri}, {Jarrett}, {Kirkpatrick}, {Padgett}, {McMillan}, {Skrutskie},
  {Stanford}, {Cohen}, {Walker}, {Mather}, {Leisawitz}, {Gautier}, {McLean},
  {Benford}, {Lonsdale}, {Blain}, {Mendez}, {Irace}, {Duval}, {Liu}, {Royer},
  {Heinrichsen}, {Howard}, {Shannon}, {Kendall}, {Walsh}, {Larsen}, {Cardon},
  {Schick}, {Schwalm}, {Abid}, {Fabinsky}, {Naes}, \& {Tsai}}]{Wright2010}
{Wright}, E.~L., {Eisenhardt}, P. R.~M., {Mainzer}, A.~K., {et~al.} 2010, \aj,
  140, 1868

\bibitem[{{Xie} {et~al.}(2021){Xie}, {Ho}, {Zhuang}, \& {Shangguan}}]{Xie2021}
{Xie}, Y., {Ho}, L.~C., {Zhuang}, M.-Y., \& {Shangguan}, J. 2021, \apj, 910,
  124

\bibitem[{{Xu} {et~al.}(2022){Xu}, {Pinto}, {Kara}, {Masterson}, {Garc{\'\i}a},
  {Fabian}, {Parker}, {Barret}, {Alston}, \& {Cusumano}}]{Xu2022}
{Xu}, Y., {Pinto}, C., {Kara}, E., {et~al.} 2022, \mnras, 513, 1910

\bibitem[{{Zhang} \& {Lu}(2019)}]{Zhang2019}
{Zhang}, X. \& {Lu}, Y. 2019, \apj, 873, 101

\bibitem[{{Zinn} {et~al.}(2013){Zinn}, {Middelberg}, {Norris}, \&
  {Dettmar}}]{Zinn2013}
{Zinn}, P.~C., {Middelberg}, E., {Norris}, R.~P., \& {Dettmar}, R.~J. 2013,
  \apj, 774, 66

\bibitem[{{Zubovas} {et~al.}(2013){Zubovas}, {Nayakshin}, {King}, \&
  {Wilkinson}}]{Zubovas2013}
{Zubovas}, K., {Nayakshin}, S., {King}, A., \& {Wilkinson}, M. 2013, \mnras,
  433, 3079

\end{thebibliography}

%\begin{thebibliography}{73}
%\expandafter\ifx\csname natexlab\endcsname\relax\def\natexlab#1{#1}\fi
%\end{thebibliography}

\begin{appendix}

\onecolumn
\section{The sample of supermassive black holes.}
\begin{longtable}{llllllllllllllllrrr}
	\caption{The sample of supermassive black holes.}\\
%	\label{ltapp}
   \hline\hline
	Name &  redshift   &   $\log L_{\rm IR}$    &  $\log$ SFR  & $\log M_{*}/M_{\odot}$  & $\log M_{\rm BH}/M_{\odot}$ & $a$ & Mass/spin Reference \\
	(1) & (2) & (3) & (4) & (5) & (6) & (7) & (8)\\
	\hline
	\endfirsthead
	\caption{continued.}\\
	\hline
	Name &  redshift   &   $\log L_{\rm IR}$    &  $\log$ SFR  & $\log M_{*}/M_{\odot}$  & $\log M_{\rm BH}/M_{\odot}$ & $a$ & Mass/spin Reference \\
	(1) & (2) & (3) & (4) & (5) & (6) & (7) & (8)\\
	\hline
	\endhead
	\hline
	\endfoot
Mrk 359	&	0.0168	&	43.846	&	0.499	&	9.677$\pm$0.03	&	6.041	&	0.66$_{-0.54}^{+0.30}$	&	Bambi21	\\
Ark564	&	0.0247	&	43.836	&	0.489	&	8.842$\pm$0.06	&	6.041	&	0.96$_{-0.06}^{+0.01}$	&	Rey21	\\
Mrk 766	&	0.01288	&	43.999	&	0.653	&	8.031$\pm$0.04	&	6.54$_{-0.04}^{+0.05}$	&	$>$0.92	&	Bambi21	\\
NGC 4051	&	0.00234	&	43.07	&	-0.274	&	7.913$\pm$0.03	&	6.281$_{-0.18}^{+0.18}$	&	$>$0.99	&	Rey21	\\
NGC 1365	&	0.00546	&	44.612	&	1.265	&	9.655$\pm$0.02	&	7.81$_{-0.23}^{+0.53}$	&	$>$0.97	&	Bambi21	\\
1H0707-495	&	0.041	&	43.297	&	-0.048	&	7.522$\pm$0.1	&	5.81$_{-0.04}^{+0.05}$	&	$>0.988$	&	Bambi21	\\
MCG-6-30-15	&	0.00775	&	42.548	&	-0.798	&	8.522$\pm$0.07	&	6.462$_{-0.24}^{+0.27}$	&	0.91$_{-0.07}^{+0.06}$	&	Rey21	\\
NGC5506	&	0.00608	&	43.209	&	-0.137	&	7.022$\pm$0.04	&	6.70$_{-0.10}^{+0.19}$	&	0.93$_{-0.04}^{+0.04}$	&	Bambi21	\\
IRAS 13224-3809	&	0.0658	&	45.127	&	1.78	&	9.411$\pm$0.02	&	5.87$_{-0.14}^{+0.12}$	&	$>$0.975	&	Bambi21	\\
TON S180	&	0.062	&	44.056	&	0.709	&	9.46$\pm$0.08	&	6.908	&	$>0.98$	&	Rey21	\\
ESO 362-G18	&	0.0124	&	43.559	&	0.212	&	9.251$\pm$0.03	&	7.097$_{-0.16}^{+0.16}$	&	$>$0.92	&	Rey21	\\
Swift J2127.4+5654	&	0.0147	&	44.034	&	0.687	&	9.601$\pm$0.02	&	7.18$_{-0.09}^{+0.09}$	&	0.58$_{-0.17}^{+0.11}$	&	Bambi21	\\
Mrk 335	&	0.02578	&	43.716	&	0.369	&	9.669$\pm$0.05	&	7.15$_{-0.11}^{+0.11}$	&	0.99$_{-0.03}^{+0.01}$	&	Bambi21	\\
Mrk 110	&	0.0353	&	43.3	&	-0.046	&	9.22$\pm$0.11	&	7.4$_{-0.11}^{0.11}$	&	$>$0.99	&	Rey21	\\
NGC 3783	&	0.00973	&	43.725	&	0.378	&	9.2$\pm$0.02	&	7.474$_{-0.08}^{+0.08}$	&	$>0.88$	&	Rey21	\\
1H0323+342	&	0.0629	&	45.156	&	1.809	&	10.017$\pm$0.06	&	7.531$_{-0.08}^{+0.12}$	&	$>$0.9	&	Rey21	\\
NGC 4151	&	0.0033	&	42.78	&	-0.566	&	9.421$\pm$0.06	&	7.12$_{-0.15}^{+0.15}$	&	0.98$_{-0.01}^{+0.01}$	&	Bambi21	\\
Mrk  79	&	0.022	&	44.192	&	0.846	&	9.957$\pm$0.02	&	7.72$_{-0.12}^{+0.12}$	&	$>0.5$	&	Bambi21	\\
PG1229+204	&	0.0636	&	44.19	&	0.843	&	10.045$\pm$0.06	&	7.756$_{-0.22}^{+0.22}$	&	0.93$_{-0.02}^{+0.06}$	&	Rey21	\\
IRAS13197-1627	&	0.01654	&	44.419	&	1.072	&	9.011$\pm$0.02	&	7.806	&	$>$0.7	&	Rey21	\\
3C120	&	0.033	&	44.55	&	1.207	&	10.08$\pm$0.02	&	7.839$_{-0.15}^{+0.20}$	&	$>0.95$	&	Rey21	\\
Mrk 841	&	0.03642	&	43.996	&	0.649	&	9.355$\pm$0.07	&	7.897$_{-0.01}^{+0.01}$	&	$>$0.52	&	Bambi21	\\
IRAS09149-6206	&	0.0573	&	44.76	&	1.413	&	10.824$\pm$0.06	&	8.00$_{-0.3}^{+1.3}$	&	0.94$_{-0.07}^{+0.02}$	&	Bambi21	\\
Ark120	&	0.0327	&	44.196	&	0.85	&	10.464$\pm$0.04	&	8.176$_{-0.06}^{+0.06}$	&	0.83$_{-0.03}^{+0.05}$	&	Bambi21	\\
RBS1124	&	0.208	&	43.714	&	0.367	&	8.5$\pm$0.31	&	8.26$_{-0.10}^{+0.10}$	&	$>$0.97	&	Bambi21	\\
1RXS J113155.4-123155	&	0.654	&	47	&	3.657	&	11.73$\pm$0.02	&	8.30$_{-0.09}^{+0.09}$	&	0.87$_{-0.15}^{+0.08}$	&	Bambi21	\\
Fairall 9	&	0.046	&	44.188	&	0.841	&	9.151$\pm$0.07	&	8.407$_{-0.10}^{+0.10}$	&	$>$0.997	&	Bambi21	\\
1H0419-577	&	0.104	&	44.517	&	1.171	&	9.821$\pm$0.07	&	8.11$_{-0.10}^{+0.10}$	&	$>$0.96	&	Bambi21	\\
PG0804+761	&	0.1005	&	44.839	&	1.492	&	11.095$\pm$0.02	&	8.74$_{-0.05}^{+0.05}$	&	$>$0.97	&	Rey21	\\
Q 2237+0305	&	1.695	&	48.0224	&	4.675	&	13.213$\pm$0.02	&	9.0$_{-0.09}^{+0.09}$	&	0.74$_{-0.03}^{+0.06}$	&	Bambi21	\\
PG2112+059	&	0.459	&	45.643	&	2.296	&	10.595$\pm$0.1	&	9.0$_{-0.09}^{+0.09}$	&	$>$0.86	&	Bambi21	\\
H1821+643	&	0.297	&	46.417	&	3.07	&	11.178$\pm$0.04	&	9.653$_{-0.14}^{+0.14}$	&	$>$0.4	&	Rey21	\\
IRAS 00521-7054	&	0.069	&	45.084	&	1.737	&	10.278$\pm$0.02	&	7.698$_{-0.09}^{+0.09}$	&	$>0.77$	&	Bambi21	\\
IRAS13349+2438	&	0.108	&	45.274	&	1.927	&	11.368$\pm$0.02	&	7.00$_{-0.17}^{+0.09}$	&	0.80$_{-0.5}^{+0.2}$	&	Bambi21	\\
Fairall 51	&	0.0142	&	43.585	&	0.238	&	8.845$\pm$0.06	&	8.00$_{-0.09}^{+0.09}$	&	0.80$_{-0.2}^{+0.2}$	&	Bambi21	\\
Mrk 1501	&	0.087	&	44.33	&	0.983	&	10.06$\pm$0.07	&	8.26$_{-0.06}^{+0.06}$	&	$>$0.98	&	Bambi21	\\
Mrk 1018	&	0.043	&	43.021	&	-0.325	&	6.873$\pm$0.3	&	7.81	&	0.58$_{-0.74}^{+0.36}$	&	Kos22/Vas16	\\
				Mrk 509	&	0.034	&	43.825	&	0.478	&	9.823$\pm$0.12	&	8.15$_{-0.04}^{+0.04}$	&	$>$0.993	&	Bambi21	\\
3C382	&	0.055	&	44.174	&	0.827	&	10.864$\pm$0.06	&	8.01$_{-0.05}^{+0.09}$	&	0.75$_{-0.04}^{+0.07}$	&	Lu19/Wal13	\\
PKS 0558-504	&	0.137	&	45.17	&	1.823	&	9.812$\pm$0.05	&	7.62	&	$>0.8$	&	Kos22/Wal13	\\
NGC 7469	&	0.016	&	45.14	&	1.794	&	9.59$\pm$0.04	&	6.42$_{-0.09}^{+0.06}$	&	$>$0.96	&	Lu19/Wal13	\\
PDS 456	&	0.185	&	45.154	&	1.808	&	9.884$\pm$0.02	&	8.4	&	$>0.97$	&	Wil18/Wal13	\\
UGC 6728	&	0.0065	&	42.758	&	-0.588	&	7.951$\pm$0.05	&	5.55$_{-0.25}^{+0.22}$	&	$>$0.95	&	Lu19/Wal13	\\
1H1934-063	&	0.01	&	43.778	&	0.431	&	9.223$\pm$0.02	&	6.477	&	$>$0.45	&	Xu22/Jia19	\\
Mrk 279	&	0.03	&	44.293	&	0.946	&	10.433$\pm$0.02	&	7.97$_{-0.12}^{+0.09}$	&	$>$0.95	&	Lu19/Jia19	\\
Mrk 590	&	0.0263	&	44.058	&	0.711	&	10.492$\pm$0.02	&	7.55$_{-0.07}^{+0.05}$	&	$>$0.1	&	Lu19/Jia19	\\
NGC 4748	&	0.0146	&	43.677	&	0.331	&	8.904$\pm$0.03	&	6.61$_{-0.23}^{+0.11}$	&	$>$0.8	&	Lu19/Jia19	\\
PG 0844+349	&	0.064	&	43.308	&	-0.037	&	7.045$\pm$0.28	&	8.35$_{-0.26}^{+0.26}$	&	$>$0.95	&	Afa19/Jia19	\\       	   
\end{longtable}
\tablefoot{{\bf Notes.} Columns (1) is the name of the sources;  Columns (2) is the redshift; Columns (3) is the torus-subtracted IR luminosity in the 8–1000~$\mu$m band in units erg s$^{-1}$; Columns (4) is SFR calculated from the torus-subtracted IR luminosity in the 8–1000 $\mu$m band, units M$_{\odot} yr^{-1}$; Columns (5) is stellar mass; Columns (6) is the black hole mass; Columns (7) is the black hole spin; Columns (8) is the reference of black hole mass and spin. Rey21=\cite{Reynolds2021}; Bambi21=\cite{Bambi2021}; Afa19=\cite{Afanasiev2019}; Kos22=\cite{Koss2022}; Vas16=\cite{Vasudevan2016}; Wal13=\cite{Walton2013}; Lu19=\cite{Lu2019}; Xu22=\cite{Xu2022}; Jia19=\cite{Jiang2019}. Some of the sources in the samples of \cite{Reynolds2021} were updated from \cite{Bambi2021}.}  
\end{appendix}

\clearpage
\end{document}